\documentclass[a4paper,12pt]{article}
\usepackage[top=1.25in, bottom=1.25in, left=1.05in, right=1.05in]{geometry}
\usepackage[utf8]{inputenc}
\usepackage[T1]{fontenc}
\usepackage[english]{babel}

\usepackage{natbib}
\usepackage[section]{placeins}
\usepackage{setspace}
\usepackage{enumitem}
\setlist{noitemsep}  

\usepackage{regexpatch}

\usepackage{amscd,amsmath,amssymb,amsfonts,amsthm,bbm,bm,latexsym,mathrsfs,mathtools,bigints}
\usepackage[subnum]{cases}	
\usepackage{dsfont}
\makeatletter
\newcommand*{\distas}[1]{\mathbin{\overset{#1}{\kern\z@\sim}}}	
\makeatother
\newcommand*\abs[1]{\left|#1\right|}		
\newcommand{\norm}[1]{\left\lVert#1\right\rVert} 		

\allowdisplaybreaks

\newtheorem{definition}{Definition}[section]
\theoremstyle{remark}

\theoremstyle{plain}
\newtheorem{exa}{Example}

\usepackage{algorithm, algorithmic}

\RequirePackage[colorlinks=true, citecolor=blue, urlcolor=blue]{hyperref}

\usepackage{xcolor,epsfig,epstopdf,rotating,graphics,graphicx, subcaption} 
\usepackage[width=0.9\textwidth, font={footnotesize}]{caption}

\usepackage{rotating,lscape,pdflscape}
\usepackage{booktabs,longtable,float,array,multirow,colortbl,adjustbox}
\newcolumntype{C}[1]{>{\centering\arraybackslash}p{#1}}
\definecolor{dgray}{gray}{0.65}
\definecolor{lgray}{gray}{0.80}
\definecolor{bostonred}{rgb}{0.8, 0.0, 0.0}
\definecolor{candyapplered}{rgb}{1.0, 0.03, 0.0}
\definecolor{ferrarired}{rgb}{1.0, 0.11, 0.0}
\colorlet{lred}{candyapplered!30!white}
\colorlet{dred}{ferrarired!50!lred}
\colorlet{dred}{dgray!90!white}
\colorlet{lred}{lgray!40!white}

\usepackage{pifont}
\newcommand{\cmark}{\text{\ding{51}}}
\newcommand{\xmark}{\text{\ding{55}}}

\usepackage{physics}
\usepackage{comment}

\usepackage{epsfig}

\def\I {\mathbb{I}}
\def\Id {\mathbf{I}}
\def\ba {\mathbf{a}}
\def\bb {\mathbf{b}}
\def\bw {\mathbf{w}}
\def\bA {\mathbf{A}}
\def\bB {\mathbf{B}}
\def \R {\mathds{R}}
\def\bpsi {\boldsymbol{\psi}}
\def\btau {\boldsymbol{\tau}}
\def\bphi {\boldsymbol{\phi}}
\def\balpha {\boldsymbol{\alpha}}
\def\PIP {\text{PIP}}
\def\RI {\text{RI}}
\def\tr {\mathrm{tr}}
\def\rmax {r_{\max}}
\def\F {\text{F}}
\def\MCC {\text{MCC}}
\def\TP {\text{TP}}
\def\TN {\text{TN}}
\def\FP {\text{FP}}
\def\FN {\text{FN}}

\title{\vspace{-60pt} \textbf{Uncertainty Quantification in Bayesian Reduced-Rank Sparse Regressions}
}

\author{
Maria F. Pintado\thanks{Queen Mary University of London, United Kingdom,  {\color{blue}\texttt{m.f.pintadoserrano@qmul.ac.uk}}, supported by CONAHCyT (Mexico), grant no. 2021-000007-01EXTF-00090.}
\and
Matteo Iacopini\thanks{Queen Mary University of London, United Kingdom, \color{blue}\texttt{m.iacopini@qmul.ac.uk}}
\and
Luca Rossini\thanks{University of Milan, Italy and Fondazione Eni Enrico Mattei, \color{blue}\texttt{luca.rossini@unimi.it}}
\and
Alexander Y. Shestopaloff\thanks{Queen Mary University of London, United Kingdom and Memorial University of Newfoundland, Canada, \color{blue}\texttt{a.shestopaloff@qmul.ac.uk}}
}

\date{\today}

\begin{document}

\maketitle

\begin{abstract}
Reduced-rank regression recognises the possibility of a rank-deficient matrix of coefficients.  We propose a novel Bayesian model for estimating the rank of the coefficient matrix, which obviates the need for post-processing steps and allows for uncertainty quantification.  Our method employs a mixture prior on the regression coefficient matrix along with a global-local shrinkage prior on its low-rank decomposition.  Then, we rely on the Signal Adaptive Variable Selector to perform sparsification and define two novel tools: the Posterior Inclusion Probability uncertainty index and the Relevance Index. The validity of the method is assessed in a simulation study, and then its advantages and usefulness are shown in real-data applications on the chemical composition of tobacco and on the photometry of galaxies. 

 \vspace*{3pt}
\end{abstract}

\section{Introduction} \label{sec:introduction}

A common problem found in several fields ranging from economics and finance to biology and medicine is the need to study the relationship between response variables and their predictors.  For instance, biochemical data from a study by \citet{Smith1962} was modelled using multivariate linear regression \citep{Reinsel2022book} to study the influence of certain characteristics of urine specimens over others under the context of reduced-rank regression.  \citet{Joreskog1975} applied a similar model in sociology considering unobservable latent variables.  \citet{Gudmundsson1977} constructed linear combinations of variables of an econometric model of the United Kingdom using time series to represent aspects of the economic situation. 

Multivariate linear regression is an extensively used model that provides a straightforward interpretation of the relationship between a group of responses and a common set of predictors.  This model is particularly useful when dealing with data where there are multiple dependent variables, as it allows for the analysis of the joint effect of the predictors on these variables.  

Reduced-rank regression (RR) exploits the dependence structure among the responses and allows the regression coefficient matrix $C$ to be rank deficient, implying a reduction in the number of parameters through linear restrictions on the entries of $C$.  It considers a reduced-rank decomposition of the coefficient matrix into the product $BA'$, with $A\in\mathbb{R}^{q\times r}$ and $B\in\mathbb{R}^{p\times r}$ dependent on the rank $r$ of $C$,
The first approach to this method was made by \citet{Anderson1951}, who proposed a class of regression models that restrict $C$ to be rank deficient.  
\citet{Izenman1975} introduced the term reduced rank regression for these models and provided a further study of the estimators.

A closely related topic of research is low-rank matrix completion, where the aim is to complete the missing entries of a matrix from a partial observation \citep[see][for a revision of Bayesian methods]{Alquier2013}. The similarity with reduced-rank regression lies in the estimation a low-rank matrix, where the rank is user-defined \citep{Lim2007,Salakhutdinov2008} or implicitly estimated \citep{babacan2011low}. The previous authors assign Gaussian priors with mean zero on $A$ and $B$, which is similar to our method. This is in contrast the proposed entry-wise uniform prior of \cite{mai2015bayesian}, which can similarly be used within our proposed mixture approach.
%

From a Bayesian perspective, \citet{Geweke1996} pioneered the early work on reduced-rank regression by assigning independent Gaussian priors on the elements of the coefficient matrix conditioning on the rank, assumed to be known.  In addition, Geweke proposed to use predictive odds ratios for regression models with different ranks when $r$ is unknown to identify a true model.  
%
Although further methods treating the rank, $r$, as unknown have been developed, they typically treat it as a parameter to be fixed before performing the inference \citep{Chen2012,Goh2017}.  The literature that treats the rank as an unknown quantity relies on post-processing steps to estimate it, for example, by thresholding the singular values \citep{Bunea2011,Chen2013,Chakraborty2019}.  The performance of post-processing methods typically depends on some user-specified tuning parameters, whose choice is hardly justifiable; moreover, it does not allow uncertainty quantification. 
The literature on nonparametric reduced-rank regression treats the rank as a tuning parameter chosen by cross-validation \citep{lian2013reducedrank}, estimates it by thresholding singular values along with parameters to be selected by the user \citep{Mukherjee2013} or by imposing regularization penalties \citep{foygel2012nonparametric}. Additionally, the nonparametric techniques equally fail to fully incorporate the quantification of uncertainty.
To address these issues, we propose a Bayesian Rank Estimation and Covariate Selection (BRECS) method, with the crucial difference that estimation of the rank is done online, jointly with all the other parameters, in a fully Bayesian approach.  As such, the proposed method allows for uncertainty quantification and removes the need for a post-processing scheme.  This is done by assuming a finite mixture prior to the coefficient matrix conditioning on the possible values of the rank.
 
Sparsity-inducing estimators of the coefficient matrix have been used to overcome the over-parametrization and potential over-fitting that typically characterise high-dimensional models.  Therefore, several authors have expanded Bayesian methods for incorporating sparsity into reduced rank regression \citep{Zhu2014,Goh2017,Chakraborty2019,Yang2022}. 
We impose a global-local shrinkage prior on the columns of $B$, which encourages sparsity in the matrix of coefficients \citep{Bhattacharya2015}.

\begin{table}[t!h]
\centering
\renewcommand{\tabcolsep}{0.27cm}
\begin{tabular}{@{} l c c c c c c @{}}
    \toprule
    \multirow{2}{*}{Model} & 
    \multicolumn{1}{p{1.5cm}}{\centering Dimension reduction} & 
    \multicolumn{1}{p{1.6cm}}{\centering Rank \\ estimation} & 
    \multicolumn{1}{p{1.6cm}}{\centering Sparse \\ estimation} & 
    \multicolumn{1}{p{1.6cm}}{\centering Unc.Quant. \\ on rank} & 
    \multicolumn{1}{p{1.7cm}}{\centering Unc.Quant. \\ on sparsity} \\
    \midrule
    Linear        & \xmark & \xmark & \xmark & \xmark & \xmark \\
    Linear Sparse & \xmark & \xmark & \cmark & \xmark & \xmark \\
    Standard RR   & \cmark & \cmark & \xmark & \xmark & \xmark \\
    BRECS          & \cmark & \cmark & \cmark & \cmark & \cmark \\ 
    \bottomrule
\end{tabular}
\caption{Comparison of linear models in terms of reduced rank and sparse estimation.}
\label{tab:parameters}
\end{table}

Moreover, the above-mentioned linear associations are likely to involve a small subset of the explanatory variables, leading to a sparse coefficient matrix \citep{Goh2017}.
Global-local shrinkage priors ensure that the coefficients are pulled towards zero, but exact sparsification (i.e., estimated coefficients exactly equal to zero) is prevented by the continuity of the prior, thus the need for an additional step for coefficient selection. Consequently, the uncertainty about this mechanism also becomes relevant.  
%
%
\citet{ray2018signal} proposed the Signal Adaptive Variable Selector (SAVS) to post-process a point estimate, such as the posterior mean, and group coefficients into exact zeros and non-zeros.  \citet{huber2021savs} applied the SAVS to each MCMC draw of the parameter of interest, thus obtaining a posterior inclusion probability (PIP) for each coefficient.  We adopt a similar strategy and apply SAVS online to each element of the coefficient matrix $C$; then, we derive a new PIP uncertainty index to quantify uncertainty in coefficient selection (see Table~\ref{tab:parameters}). A related work by \cite{yuchi2023bayesian} provides another way of defining a prior (conditional on the rank) and different uncertainty quantification measures for a low-rank matrix in the context of matrix completion.

In addition to the PIP uncertainty index, we also introduce the Relevance Index. The PIP uncertainty index allows us to quantify uncertainty about variable inclusion. The Relevance Index, which is the most important one, is a distribution representing the relevance of a covariate in terms of the share of response variables on which it has a significant impact. This index provides full uncertainty quantification.  Moreover, we propose a rule of thumb for variable selection that summarises and complements the information embedded in the relevance index by means of its survival function.

Uncertainty quantification in both rank estimation and variable selection is currently underdeveloped in the literature. \cite{Yang2022} address this issue by using the Laplace approximation of the posterior distributions within a collapsed Gibbs sampler and obtaining complete sparse rows of $C$ for covariate selection.  By coupling our mixture prior on $C$  with the shrinkage prior on $B$, our proposed method enables sparse and low-rank estimation of the coefficient matrix by sampling from the exact full-conditional posteriors, obviating the demand for an approximation. Our approach removes the need for post-processing while jointly incorporating the quantification of uncertainty. Our approach, different from the literature, does not rely on visual inspection of the plots and on user-specified thresholds to choose the rank. However, we allow for a simple and transparent interpretation based on the entire posterior distribution of the rank. Practically being able to quantify the uncertainty in the rank allows a user to assess how well the data guides in choosing a specific rank value in a concrete setting.
Besides, differently from \cite{Yang2022}, our approach is able to obtain a sparse estimate of $C$ where either entire rows or only single entries are null. The PIP reports uncertainty quantification on the estimates of the rank and the coefficients, and then the Relevance Index (RI) is used to measure the uncertainty about variable selection.

Simulation studies were conducted to evaluate the performance of the proposed methodology in various scenarios, and compared to other relevant methods.
The effectiveness of the algorithm is validated as well in real data application to datasets on the chemical composition of tobacco and on the photometry of galaxies.  For the tobacco dataset, we obtain comparable results with \cite{Izenman2008book} while requiring the estimation of a single model and uncertainty quantification without relying on a subjective judgment. For the Galaxy experiment, we provide a strong and interesting variable selection. Furthermore, the proposed relevance index and its survival function allow us to select covariates with different degrees of uncertainty, which is not possible with the commonly used post-processing methods.

The remainder of the article is organised as follows.
Section~\ref{sec:model} introduces the model, and presents the proposed priors for rank selection.  Then, Section~\ref{sec:posterior} demonstrates our sampling algorithm and provides different definitions of uncertainty quantification.  Section~\ref{sec:simulations} illustrates the performance of the proposed methods in simulated experiments, while Section~\ref{sec:application} applies them to two real datasets.  Finally, Section~\ref{sec:conclusion} provides a discussion. 

\section{Reduced-rank regression model} \label{sec:model}

For each observational unit $i=1,\ldots,n$ from a sample of size $n$, let $\boldsymbol y_i \in \mathbb{R}^q$ be a response variable explained by $p$ possible predictors $\boldsymbol x_i \in \mathbb{R}^p$.  Let $Y \in \mathbb{R}^{n\times q}$ be the matrix of responses with the $i$th row as $\boldsymbol y_i'$, and $X\in\mathbb{R}^{n\times p}$ the matrix of explanatory variables with the $i$th row as $\boldsymbol x_i'$.  The multivariate linear regression model is defined as
\begin{equation}\label{eq:linearmodel}
    Y=XC+E, \qquad E=(\boldsymbol e_1,\ldots,\boldsymbol e_n)',
\end{equation}
where the rows $\boldsymbol e_i$ of the error matrix are independent and normally distributed with mean zero and $q\times q$ covariance matrix $\Sigma$.  

Reduced-rank regression identifies a smaller set of variables that can explain a large proportion of the variation in the data.  To consider this model, an assumption on the rank of the matrix of coefficients $C$ is defined as rank$(C)=r \leq \min(p,q)$.  Such an assumption translates into fewer parameters, leading to a more parsimonious model.  In the literature, the constraint on the rank has been done by assuming that $r$ is known \citep{Geweke1996}, or declaring it as unknown but fixed as in \citet{Chen2012} and \citet{Goh2017}.  Typically, there is no guidance in fixing a specific rank for $C$ in real data applications, motivating the choice of treating it as an unknown quantity.  However, previous results following this approach rely on post-processing schemes to estimate the rank with user-tuned parameters \citep{Chakraborty2019,mai2021efficient}.  In the proceeding of the paper, we consider the low-rank decomposition $C=BA'$ with $B \in \mathbb{R}^{p\times r}$ and $A \in \mathbb{R}^{q\times r}$, where the rank $r$ needs to be estimated.

\subsection{Mixture prior for rank selection} \label{sec:prior}

In this paper, we contribute to the literature on reduced rank regression by proposing a new Bayesian approach for rank estimation and related uncertainty quantification. Our proposal relies on finite mixture priors, which combine 
two or more prior probability distributions, namely the mixture components, each with its own set of parameters.  The main aspect that favours a mixture prior is that rank selection, which corresponds to an automatic model choice, is made along with parameter estimation.  In the proposed mixture prior, each component corresponds to a different rank.  The Bayesian approach to inference coupled with data augmentation allows for obtaining a posterior distribution for the rank.  Besides deriving a point estimate as the maximum a posteriori (MAP), our approach permits uncertainty quantification, a novel feature of this method in contrast to existing literature.  

In detail, we define a finite mixture prior on $C$ made by a number of components equal to $\rmax=\min(p,q)$, assumed to be $q$. Employing the notation $C_s$ for the matrix $C$ under the restriction $\operatorname{rank}(C)=s$, the prior is expressed as
\begin{equation}
\label{eq:priorCmix}
    p(C) = \sum_{s=1}^q w_s p(C_s),
\end{equation}
where $w_s$ is the prior probability of rank$(C)=s$.  Denoting $\bw = (w_1,\ldots,w_q)$, we assume a Dirichlet prior distribution with parameter $\boldsymbol\gamma = (\gamma_1,\ldots,\gamma_q)$, that is $\bw \sim \mathcal{D}ir(\boldsymbol\gamma)$.
We introduce a latent allocation variable $u$ which assumes values in the set $\left\{1,\ldots,q\right\}$, and follows a prior Categorical distribution with parameter $\bw$, represented as
\begin{equation}\label{eq:prioru}
    p(u|\bw)=\prod_{s=1}^q w_s^{\I(u=s)},
\end{equation}    
where $\I(u=s)$ is the indicator function, taking value $1$ if $u=s$ and $0$ otherwise.

To define a prior on $C$ conditional on its rank $u$,  we rely on the low-rank representation $C=BA'$, and assume prior independence between $A$ and $B$, which results in
\begin{equation}\label{eq:priorCs}
    p(C_u)= p(A_u)p(B_u),
\end{equation}
where $A_u$ and $B_u$ are the factors of the rank-$u$ decomposition of $C$, reminding that their dimensions are rank dependent, being $q\times u$ and $p\times u$, respectively.
For the entries of the matrix $A_u$, we consider a standard normal prior $a_{jh} \sim \mathcal{N}(0,1)$, with $j=1,\ldots,q$ and $h=1,\ldots,u$.
%
Regarding the prior specification for $B_u$, we use a global-local shrinkage prior on each column $\boldsymbol b_h=(b_{1h},b_{2h},\ldots,b_{ph})' \in \mathbb{R}^p$ of $B_u$, for $h=1,\ldots,u$. This family of distributions consists of a hierarchical scale mixture of (multivariate) Gaussian distributions of the type
\begin{equation}
    b_{lh} | \tau_h, \phi_{lh}  \sim \mathcal{N}(0, \tau_h \phi_{lh}), \qquad
    \tau_h  \sim \pi_{\tau_h}(\tau_h), \qquad \phi_{lh}  \sim \pi_{\phi_{lh}}(\phi_{lh}),
\end{equation}
where $\tau_h$ and $\phi_{lh}$ are the global and local components of the variance, respectively, with distributions $\pi_{\tau_h}(\cdot)$ and $\pi_{\phi_{lh}}(\cdot)$.\footnote{By specifying the distributions of the two variance components and eventually introducing additional layers in the hierarchy, it is possible to generate a wide range of prior distributions that shrink the coefficients toward zero while allowing the data to inform about large deviations from the origin thanks to the heavy tails of the marginal prior (obtained integrating out the global component $\tau$).}
In this article, we consider a Dirichlet-Laplace prior \citep{Bhattacharya2015,Cross2020}, which can be represented as
\begin{equation}
\begin{split}
    b_{lh}|\psi_{lh},\tau_h,\phi_{lh} & \sim \mathcal{N}(0,\psi_{lh}\tau_h^2\phi_{lh}^2), \quad l=1,\ldots,p\\
    \tau_h|\alpha_h & \sim \mathcal{G}a(\alpha_h p,1/2),\\
    \boldsymbol\phi_h|\alpha_h & \sim \mathcal{D}ir(\alpha_h,\ldots,\alpha_h), \\
    \psi_{lh} & \sim \mathcal{E}xp(1/2), \\
    \alpha_h & \sim \mathcal{U}(L_\alpha,U_\alpha),
\end{split}
\label{eq:priorB}
\end{equation}
where $\mathcal{G}a(\cdot)$ and $\mathcal{D}ir(\cdot)$ denote the Gamma (with the shape-rate parametrization) and Dirichlet distributions, respectively.
As usual with hierarchical priors, the performance of the DL prior depends on the hyperparameter values, particularly on $\alpha_h$. To address this issue, similarly to \cite{Cross2020}, we assume a continuous uniform prior for $\alpha_h$ to let data inform about the degree of shrinkage.
For each entry $l$ in column $h$, the local shrinkage parameter is $\psi_{lh}$, and the vector of global shrinkage is $(\tau_h\phi_{1h}, \ldots, \tau_h\phi_{ph})$, where $\boldsymbol\phi_h = (\phi_{1h}, \ldots, \phi_{ph})'$ is constrained to lie in the $(p-1)$ simplex $\Delta^{p-1}$.
Finally, we impose no restrictions on the covariance matrix $\Sigma$, and place an inverse Wishart prior, $\Sigma \sim \mathcal{IW}(\nu,\Upsilon)$.

As a consequence of the mixture prior on the matrix of coefficients $C$, the observed likelihood function is a mixture distribution with the same weights. Denoting $\bA=\{A_1,\ldots,A_q\}$ and $\bB=\{B_1,\ldots,B_q\}$, the likelihood is represented as
\begin{equation}
p(Y|\bA,\bB,\Sigma,\bw) = \sum_{s=1}^q w_s (2\pi)^{-\frac{n}{2}} |\Sigma|^{-\frac{n}{2}} \exp\Big\{ -\frac{1}{2} \tr[\Sigma^{-1}(Y-XB_{s}A_{s}')' (Y-XB_{s}A_{s}')] \Big\}
\end{equation}
and the complete data likelihood is
\begin{align}
   p(&Y,u|\bA,\bB,\Sigma,\bw) = p(Y|\bA,\bB,\Sigma,u,\bw) p(u|\bw) \\ \notag
   & = \prod_{s=1}^q \left( \frac{1}{(2\pi)^{nq/2}|\Sigma|^{n/2}} \exp{-\frac{1}{2} \tr[\Sigma^{-1}(Y-XB_{s}A_{s}')' (Y-XB_{s}A_{s}')]} w_s\right)^{\I(u=s)}.
\end{align}

\subsection{Alternative prior parametrizations} \label{sec:parametrizations}

The previous section has introduced a generic prior structure for the matrices $A_u$ and $B_u$, conditional on the rank of $C$ being $u$. However, no direct connection was assumed between $A_u$ and $A_v$, for $u \neq v$ (similarly for $B_u$). In this section, we leverage on the particular structure imposed by the reduced-rank assumption for $C$ to define two alternative prior parametrizations for the matrices $A_u$ and $B_u$.
In particular, we propose two main parametrizations for $A_u$ and $B_u$: the na\"ive (RRn) and the column-sharing (RRcs).
The first case provides the best unconditional approximation of the matrix of coefficients with the estimated rank, a parametrization that results in a computationally intensive algorithm with $O(q^3+q^2p)$ parameters.
In contrast, RRcs is based on sharing information across columns, leading to the best conditional approximation, a reduced number of parameters of $O(q^2+pq)$, and a computationally faster MCMC for posterior inference compared to the former approach.

\begin{figure}[t!h]
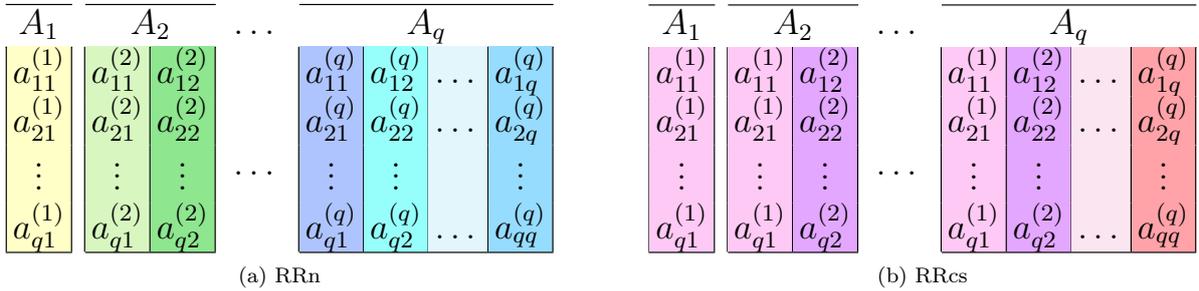

\centering
\renewcommand{\tabcolsep}{2pt}
\setlength{\abovecaptionskip}{6pt}
\begin{subfigure}{0.46\linewidth}
    \setlength{\abovecaptionskip}{5pt}
    \resizebox{\textwidth}{!}{%
    \begin{tabular}{ccc ccc ccc cc}  
      \cline{1-1} \cline{3-4} \cline{8-11}
      $A_1$ &
       &
      \multicolumn{2}{c}{$A_2$} &
       & \ldots &
       &
      \multicolumn{4}{c}{$A_q$} \\ \cline{1-1} \cline{3-4} \cline{8-11}
     
      \multicolumn{1}{|c|}{\cellcolor[HTML]{FFFFC7}$a_{11}^{(1)}$} &
      \multicolumn{1}{c|}{} &
      \multicolumn{1}{c|}{\cellcolor[HTML]{D8F7C0}$a_{11}^{(2)}$} &
      \multicolumn{1}{c|}{\cellcolor[HTML]{8EE68F}$a_{12}^{(2)}$} &
      \multicolumn{1}{c}{} &
      \multicolumn{1}{c}{} &
      \multicolumn{1}{c|}{} &
      \multicolumn{1}{c|}{\cellcolor[HTML]{AEC4FE}$a_{11}^{(q)}$} &
      \multicolumn{1}{c|}{\cellcolor[HTML]{96FFFB}$a_{12}^{(q)}$} &
      \multicolumn{1}{c|}{\cellcolor[HTML]{E2F6FC}\ldots} &
      \multicolumn{1}{c|}{\cellcolor[HTML]{96DCFE}$a_{1q}^{(q)}$} \\
     
      \multicolumn{1}{|c|}{\cellcolor[HTML]{FFFFC7}$a_{21}^{(1)}$} &
      \multicolumn{1}{c|}{} &
      \multicolumn{1}{c|}{\cellcolor[HTML]{D8F7C0}$a_{21}^{(2)}$} &
      \multicolumn{1}{c|}{\cellcolor[HTML]{8EE68F}$a_{22}^{(2)}$} &
      \multicolumn{1}{c}{} &
      \multicolumn{1}{c}{} &
      \multicolumn{1}{c|}{} &
      \multicolumn{1}{c|}{\cellcolor[HTML]{AEC4FE}$a_{21}^{(q)}$} &
      \multicolumn{1}{c|}{\cellcolor[HTML]{96FFFB}$a_{22}^{(q)}$} &
      \multicolumn{1}{c|}{\cellcolor[HTML]{E2F6FC}\ldots} &
      \multicolumn{1}{c|}{\cellcolor[HTML]{96DCFE}$a_{2q}^{(q)}$} \\
     
      \multicolumn{1}{|c|}{\cellcolor[HTML]{FFFFC7}\vdots} &
      \multicolumn{1}{c|}{} &
      \multicolumn{1}{c|}{\cellcolor[HTML]{D8F7C0}\vdots} &
      \multicolumn{1}{c|}{\cellcolor[HTML]{8EE68F}\vdots} &
      \multicolumn{1}{c}{} &
      \multicolumn{1}{c}{} &
      \multicolumn{1}{c|}{} &
      \multicolumn{1}{c|}{\cellcolor[HTML]{AEC4FE}\vdots} &
      \multicolumn{1}{c|}{\cellcolor[HTML]{96FFFB}\vdots} &
      \multicolumn{1}{c|}{\cellcolor[HTML]{E2F6FC}} &
      \multicolumn{1}{c|}{\cellcolor[HTML]{96DCFE}\vdots} \\
    
      \multicolumn{1}{|c|}{\cellcolor[HTML]{FFFFC7}$a_{q1}^{(1)}$} &
      \multicolumn{1}{c|}{} &
      \multicolumn{1}{c|}{\cellcolor[HTML]{D8F7C0}$a_{q1}^{(2)}$} &
      \multicolumn{1}{c|}{\cellcolor[HTML]{8EE68F}$a_{q2}^{(2)}$} &
      \multicolumn{1}{c}{} &
      \multirow{-4}{*}{\ldots} &
      \multicolumn{1}{c|}{} &
      \multicolumn{1}{c|}{\cellcolor[HTML]{AEC4FE}$a_{q1}^{(q)}$} &
      \multicolumn{1}{c|}{\cellcolor[HTML]{96FFFB}$a_{q2}^{(q)}$} &
      \multicolumn{1}{c|}{\cellcolor[HTML]{E2F6FC}\ldots} &
      \multicolumn{1}{c|}{\cellcolor[HTML]{96DCFE}$a_{qq}^{(q)}$} \\ \cline{1-1} \cline{3-4} \cline{8-11}
    \end{tabular}%
    }
    \caption{RRn}
    \label{fig:RRn}
\end{subfigure}
\hfill
\begin{subfigure}{0.46\linewidth}
    \setlength{\abovecaptionskip}{5pt}
    \resizebox{\textwidth}{!}{%
    \begin{tabular}{ccc ccc ccc cc}
      \cline{1-1} \cline{3-4} \cline{8-11}
      $A_1$ &
       &
      \multicolumn{2}{c}{$A_2$} &
       & \ldots &
       &
      \multicolumn{4}{c}{$A_q$} \\ \cline{1-1} \cline{3-4} \cline{8-11}

      \multicolumn{1}{|c|}{\cellcolor[HTML]{FFC9F5}$a_{11}^{(1)}$} &
      \multicolumn{1}{c|}{} &
      \multicolumn{1}{c|}{\cellcolor[HTML]{FFC9F5}$a_{11}^{(1)}$} &
      \multicolumn{1}{c|}{\cellcolor[HTML]{E4A7FF}$a_{12}^{(2)}$} &
      \multicolumn{1}{c}{} &
      \multicolumn{1}{c}{} &
      \multicolumn{1}{c|}{} &
      \multicolumn{1}{c|}{\cellcolor[HTML]{FFC9F5}$a_{11}^{(1)}$} &
      \multicolumn{1}{c|}{\cellcolor[HTML]{E4A7FF}$a_{12}^{(2)}$} &
      \multicolumn{1}{c|}{\cellcolor[HTML]{FBE5F0}\ldots} &
      \multicolumn{1}{c|}{\cellcolor[HTML]{FFA3A9}$a_{1q}^{(q)}$} \\

      \multicolumn{1}{|c|}{\cellcolor[HTML]{FFC9F5}$a_{21}^{(1)}$} &
      \multicolumn{1}{c|}{} &
      \multicolumn{1}{c|}{\cellcolor[HTML]{FFC9F5}$a_{21}^{(1)}$} &
      \multicolumn{1}{c|}{\cellcolor[HTML]{E4A7FF}$a_{22}^{(2)}$} &
      \multicolumn{1}{c}{} &
      \multicolumn{1}{c}{} &
      \multicolumn{1}{c|}{} &
      \multicolumn{1}{c|}{\cellcolor[HTML]{FFC9F5}$a_{21}^{(1)}$} &
      \multicolumn{1}{c|}{\cellcolor[HTML]{E4A7FF}$a_{22}^{(2)}$} &
      \multicolumn{1}{c|}{\cellcolor[HTML]{FBE5F0}\ldots} &
      \multicolumn{1}{c|}{\cellcolor[HTML]{FFA3A9}$a_{2q}^{(q)}$} \\

      \multicolumn{1}{|c|}{\cellcolor[HTML]{FFC9F5}\vdots} &
      \multicolumn{1}{c|}{} &
      \multicolumn{1}{c|}{\cellcolor[HTML]{FFC9F5}\vdots} &
      \multicolumn{1}{c|}{\cellcolor[HTML]{E4A7FF}\vdots} &
      \multicolumn{1}{c}{} &
      \multicolumn{1}{c}{} &
      \multicolumn{1}{c|}{} &
      \multicolumn{1}{c|}{\cellcolor[HTML]{FFC9F5}\vdots} &
      \multicolumn{1}{c|}{\cellcolor[HTML]{E4A7FF}\vdots} &
      \multicolumn{1}{c|}{\cellcolor[HTML]{FBE5F0}} &
      \multicolumn{1}{c|}{\cellcolor[HTML]{FFA3A9}\vdots} \\

      \multicolumn{1}{|c|}{\cellcolor[HTML]{FFC9F5}$a_{q1}^{(1)}$} &
      \multicolumn{1}{c|}{} &
      \multicolumn{1}{c|}{\cellcolor[HTML]{FFC9F5}$a_{q1}^{(1)}$} &
      \multicolumn{1}{c|}{\cellcolor[HTML]{E4A7FF}$a_{q2}^{(2)}$} &
      \multicolumn{1}{c}{} &
      \multirow{-4}{*}{\ldots} &
      \multicolumn{1}{c|}{} &
      \multicolumn{1}{c|}{\cellcolor[HTML]{FFC9F5}$a_{q1}^{(1)}$} &
      \multicolumn{1}{c|}{\cellcolor[HTML]{E4A7FF}$a_{q2}^{(2)}$} &
      \multicolumn{1}{c|}{\cellcolor[HTML]{FBE5F0}\ldots} &
      \multicolumn{1}{c|}{\cellcolor[HTML]{FFA3A9}$a_{qq}^{(q)}$} \\ \cline{1-1} \cline{3-4} \cline{8-11}
    \end{tabular}%
    }
    \caption{RRcs}
    \label{fig:RRcs}
\end{subfigure}
\caption{Matrix $A_u$ under both parametrizations: the na\"ive (panel a) and the column-sharing (panel b). Each colour represents a set of values for the corresponding columns of $A_u$. In RRn, the elements of the first (and only) column of $A_1$ are different from the first column of $A_2,\ldots,A_q$. In RRcs, the elements of the first (and only) column in $A_1$ are the same as those of the first column of $A_2,\ldots,A_q$.}
\label{fig:RRnvsRRcs}
\end{figure}

The auxiliary variable $u$ represents the (unknown) rank of $C$, thus implicitly determining the number of columns of $A$ and $B$.
Within the RRn parametrization, each value of $u$ is associated with a specific collection of matrices $A_u \in\R^{q\times u}$, $B_u \in\R^{p\times u}$ and the corresponding parameters of the hierarchical prior for each column of $B_u$, that is $(\phi,\tau,\psi,\alpha)$. As $u$ ranges from $1$ to $q$, there are in total $q$ collections of parameters, also differing in the number of elements included in each collection (Figure~\ref{fig:RRn}).
Instead, in the RRcs parametrization, moving from $u$ to $u+1$ implies sharing the same parameters as in $u$, plus an additional column of the matrices $A_{u+1}$, $B_{u+1}$ (and the corresponding parameters $\phi,\tau,\psi,\alpha$). 
Therefore, the sharing mechanism conditions the reduced-rank approximation of the matrix $C_{u+1}$ to the \textit{even lower} rank approximation $C_u$ (Figure~\ref{fig:RRcs}).

To summarise, the crucial difference is the parametrization of the collection of matrices $\{ A_u, B_u \}_{u=1}^q$. Denoting with $\ba_h^{(u)}$ and $\bb_h^{(u)}$ the $h$th column of the matrices $A_u$ and $B_u$, the RRn parametrization assumes:
\begin{align}
    \nonumber
    A_1 & = [ \ba_{1}^{(1)} ]
    && B_1  = [ \bb_{1}^{(1)} ] \\
    \nonumber
    A_2 & = [ \ba_{1}^{(2)}, \ba_{2}^{(2)} ]
    && B_2  = [ \bb_{1}^{(2)}, \bb_{2}^{(2)} ] \\
     & \vdots  && & \\
    \nonumber
    A_q & = [ \ba_{1}^{(q)}, \ba_{2}^{(q)}, \ldots, \ba_{q}^{(q)} ]
    && B_q  = [ \bb_{1}^{(q)}, \bb_{2}^{(q)}, \ldots, \bb_{q}^{(q)} ].
\end{align}
Conversely, the RRcs parametrization assumes:
\begin{align}
    \nonumber
    A_1 & = [ \ba_{1}^{(1)} ]
    && B_1  = [ \bb_{1}^{(1)} ] \\
    \nonumber
    A_2 & = [ \ba_{1}^{(1)}, \ba_{2}^{(2)} ]
    && B_2  = [ \bb_{1}^{(1)}, \bb_{2}^{(2)} ] \\
     & \vdots  && & \\
    \nonumber
    A_q & = [ \ba_{1}^{(1)}, \ba_{2}^{(2)}, \ldots, \ba_{q}^{(q)} ]
    && B_q  = [ \bb_{1}^{(1)}, \bb_{2}^{(2)}, \ldots, \bb_{q}^{(q)} ]
\end{align}
The first parametrization produces a different estimate of the $h$th column of each low-rank matrix: $\ba_h^{(u)} \neq \ba_h^{(v)}$, for $u\neq v$, and $h\leq \min(u,v)$. In contrast, the second parametrization assumes that $\ba_h^{(u)} = \ba_h^{(v)}$. The same rationale applies to matrix $B$. 

Concerning the prior construction for the columns of either matrix, we assume the same distributions for the RRn and RRcs, that is $\ba_h^{(u)} \sim \mathcal{N}(\mathbf{0},\Id)$ and eq.~\eqref{eq:priorB} for each $\bb_h^{(u)}$.

\section{Posterior sampling} \label{sec:posterior}

In this section, we yield the estimation details of  the proposed mixture prior and derive the model uncertainty quantification indexes.  Initially, we provide the representation of the joint posterior distribution of the model parameters. Let us define $\Psi=(\bpsi_1=(\psi_{11},\ldots,\psi_{p1}),\ldots,\bpsi_q=(\psi_{1q},\ldots,\psi_{pq}))$, $\btau=(\tau_1,\ldots,\tau_q)$, $\Phi=(\bphi_1,\ldots,\bphi_q)$ and $\balpha=(\alpha_1,\ldots,\alpha_q)$. The parameters can be included in $\Theta = (\Sigma,\bw,\bA,\bB,\Psi,\btau,\Phi,\balpha)$, and the joint posterior distribution is given by
\begin{equation}
\begin{aligned}
    p(\Theta,u|Y) &\propto p(Y|\Theta,u)\, p(\bA| u)\, p(\bB| \Psi,\btau,\Phi,u)\, p(u|\bw) \\
    & \quad \times p(\btau|\balpha)\, p(\Phi|\balpha)\, p(\Sigma)\, p(\bw)\, p(\balpha) \, p(\Psi).
\end{aligned}
\end{equation}

The choice of the prior distributions allows for a straightforward implementation of an efficient Markov Chain Monte Carlo (MCMC) algorithm to update the parameters by sampling from the full conditional posterior distributions (see the Supplement for a detailed derivation).
The proposed Gibbs sampler for RRn and RRcs are outlined in Algorithm~\ref{alg:RRn} and Algorithm~\ref{alg:RRcs}, respectively, where $\text{GiG}(\cdot)$ and $\text{iG}(\cdot)$ denote the generalised inverse Gaussian and the inverse Gaussian distributions, and a $\ast$ superscript denotes the value of the hyper-parameters of the posterior distribution.
A comprehensive description of hyperparameter specifications can be found in the Supplement.

\begin{algorithm}
\caption{Gibbs sampler for RRn specification}
\label{alg:RRn}
\begin{algorithmic}[1]
\STATE Sample $u$ from the posterior distribution on a logarithmic scale through inverse transform sampling;
\STATE Sample $\bw$ from $\mathcal{D}ir(\boldsymbol \gamma^*)$;
\STATE Sample $\Sigma$ from $\mathcal{IW}(\nu^*,\Upsilon^*)$;
\FOR{$s=1$ \TO $\rmax$}
    \IF{$s=u$}
    \STATE Sample $A_u$ by drawing $\operatorname{vec}(A_u)$ from $\mathcal{N}_{qu}(\boldsymbol \mu_{A_u}^*,\Sigma_{A_u}^*)$;
    \STATE Sample $B_u$ by drawing $\operatorname{vec}(B_u')$ from $\mathcal{N}_{pu}(\boldsymbol \mu_{B_u}^*,\Sigma_{B_u}^*)$;
    \ELSE
        \STATE Sample $A_s$ and $B_s$ from the prior;
    \ENDIF
    \FOR{$h=1$ \TO $s$}
        \STATE Sample $\tau_h$ from $\text{GiG}(p_{\tau_h}^*, a_{\tau_h}^*,b_{\tau_h}^*)$;
        \STATE Sample $\tilde{\psi}_{lh}$ from $\text{iG}(a_{\psi_{lh}}^*,b_{\psi_{lh}}^*)$, then set $\psi_{lh} = \tilde{\psi}_{lh}^{-1}$, for each $l=1,\ldots,p$;
        \STATE Sample $T_{lh}$ from $\text{GiG}(p_{\phi_{lh}}^*, a_{\phi_{lh}}^*,b_{\phi_{lh}}^*)$, then set $\phi_{lh} = T_{lh}/\sum_{i=1}^p T_{ih}$, for each $l=1,\ldots,p$;
        \STATE Sample $\alpha_h$ from its full conditional using a griddy Gibbs sampler \citep{ritter1992facilitating}.
    \ENDFOR
\ENDFOR
\end{algorithmic}
\end{algorithm}

\begin{algorithm}
\caption{Gibbs sampler for RRcs specification}
\label{alg:RRcs}
\begin{algorithmic}[1]
\STATE Sample $u$ from the posterior distribution on a logarithmic scale through inverse transform sampling;
\STATE Sample $\bw$ from $\mathcal{D}ir(\boldsymbol \gamma^*)$;
\STATE Sample $\Sigma$ from $\mathcal{IW}(\nu^*,\Upsilon^*)$;
\FOR{$s=1$ \TO $u$}
    \STATE Sample $A_u$ by drawing each column $\boldsymbol a_s$ from $\mathcal{N}_{q}(\mu_{\boldsymbol a_s}^*,\Sigma_{\boldsymbol a_s}^*)$;
    \STATE Sample $B_u$ by drawing each column $\boldsymbol b_s$ from $\mathcal{N}_{p}(\mu_{\boldsymbol b_s}^*,\Sigma_{\boldsymbol b_s}^*)$;
\ENDFOR
\FOR{$s=u+1$ \TO $\rmax$}
    \STATE Sample columns $\boldsymbol a_s$ and $\boldsymbol b_s$ from the prior;
\ENDFOR
\FOR{$s=1$ \TO $\rmax$}
    \STATE Sample $\tau_s$ from $\text{GiG}(p_{\tau_s}^*, a_{\tau_s}^*,b_{\tau_s}^*)$;
    \STATE Sample $\tilde{\psi}_{ls}$ from $\text{iG}(a_{\psi_{ls}}^*,b_{\psi_{ls}}^*)$, then set $\psi_{ls} = \tilde{\psi}_{ls}^{-1}$, for each $l=1,\ldots,p$;
    \STATE Sample $T_{ls}$ from $\text{GiG}(p_{\phi_{ls}}^*, a_{\tau_{ls}}^*,b_{\tau_{ls}}^*)$, then set $\phi_{ls} = T_{ls}/\sum_{i=1}^p T_{is}$, for each $l=1,\ldots,p$;
    \STATE Sample $\alpha_s$ from its full conditional using a griddy Gibbs sampler \citep{ritter1992facilitating}.
\ENDFOR
\end{algorithmic}
\end{algorithm}

The algorithm of the column-sharing approach (RRcs) performs significantly faster than the na\"ive (RRn) parametrization since at each iteration of the MCMC, the number of columns updated (through the posterior distribution or the prior) for $\bA$ and $\bB$ is $\rmax$, the maximum possible rank, as opposed to $\rmax(\rmax+1)/2$ in the na\"ive approach. In addition to faster computational performance, the column-sharing case presents a better mixing around the rank estimate (see Section~\ref{sec:simulations}).

\subsection{Variable selection} \label{sec:SAVS}

Further interest is placed on obtaining exact zeros in the coefficient matrix $C$ for covariate selection, and uncertainty quantification of covariates inclusion becomes relevant.  Shrinkage priors allow for parameter estimates that are very close to zero but not exactly zero, and the nature of hierarchical shrinkage priors makes common MCMC methods for sparsification that rely on cross-validation to be computationally prohibitive.  However, \cite{ray2018signal} introduced the signal adaptive variable selector (SAVS), a simple algorithm for the sparsification step on the posterior mean of the parameter of interest.  Employing SAVS allows to have exact zeros, but uncertainty quantification remains uncovered.  \cite{huber2021savs} apply the SAVS method to sparsify every MCMC draw in the shrinkage step, thus allowing for parameter uncertainty.  By following their approach to sparsify each draw, we allow for parameter uncertainty quantification.  This yields the following estimate to obtain a sparse draw of $C_{jk}$ at the $m$th iteration of the MCMC:
\begin{align}
   \bar{C}_{jk}^{(m)} = \operatorname{sign}\big( C_{jk}^{(m)} \big) \, \norm{X_j}^{-2} \left( |C_{jk}^{(m)}| \, \norm{X_j}^{2} - |C_{jk}^{(m)}|^{-2} \right)_+
\end{align}
with $X_j = (X_{1j},\ldots,X_{nj})'$ denoting the $j$th column of the matrix $X$, $(x)_+ = \max(x,0)$ and $\operatorname{sign}(x)=1$ for $x\geq 0$ and $-1$ otherwise.
Considering an MCMC of length $M$ iterations, by applying the SAVS at each iteration $m$, we obtain a collection of $M$ sparse estimates of every element $C_{jk}$. Therefore, the posterior probability that the coefficient $C_{jk}$ is not zero is the proportion of MCMC iterations such that the estimate $\bar{C}_{jk} \neq 0$. We define this proportion as the posterior inclusion probability (PIP):
\begin{align} \label{eq:PIP}
    \PIP_{jk} = \frac{1}{M}\sum_{m=1}^M \I\left( \bar{C}_{jk}^{(m)} \neq 0 \right).
\end{align}
The posterior estimate of the $jk$th entry of $C$ is set to $0$ if $\PIP_{jk} \leq 0.5$ and to the posterior mean of $\bar{C}_{jk}$ if $\PIP_{jk} > 0.5$.

By definition, the PIP is a value between $0$ and $1$, where a PIP close to $0$ indicates that the corresponding entry is not likely to be important. In contrast, a PIP close to $1$ suggests that the inclusion of the entry is well-supported by the data.  Hence, for both high and low PIPs, the decision about including an element is less uncertain, opposite to PIPs that lie around $0.5$.  Even though PIPs are, by nature, a quantification of uncertainty, their direct interpretation may not appear evident to the reader as the degree of certainty is not monotonous.  The need for a straightforward and easily interpreted manner to quantify uncertainty about variable inclusion leads us to the following definition.

\begin{definition}[PIP uncertainty index]
Let us assume the posterior inclusion probability (PIP) as in Eq.~\eqref{eq:PIP}, the PIP uncertainty index, which takes values in $[0,1]$ is defined as:
\begin{equation}
    \zeta_{jk} = 1 - 2 \big| \PIP_{jk} - 0.5 \big|,
\end{equation}
where $\zeta_{jk}$ close to $0$ (to $1$) means low (high) uncertainty about the decision of setting the $jk$th entry of $C$ to an exact $0$ or not.\footnote{The quantity $\zeta_{jk}$ is in strategy comparable to the Bernoulli variance. The linear transformation of $\PIP_{jk}$ allows equal weights for all probabilities, in contrast to different lengths in the range of values, pointing to different levels of variation in the Bernoulli variance.  In this sense, $\zeta_{jk}$ facilitates the interpretability of the results. A sensitivity analysis about the choice of $\zeta_{jk}$ is included in the Supplement.}
\end{definition}

The following example illustrates the previous definition.
\begin{exa}
Let us consider three different entries: $jk$, $jk^*$ and $jk^{**}$. We assume $\PIP_{jk} = 0.49$, $\PIP_{jk^*} = 0.94$, and $\PIP_{jk^{**}} = 0.01$.  Then, the point estimates are $\hat{C}_{jk} = \hat{C}_{jk^{**}} = 0$, whereas $\hat{C}_{jk^*} = M^{-1} \sum_{m=1}^M \bar{C}_{jk*}^{(m)}$.  The associated uncertainty indices are $\zeta_{jk}=0.98$, $\zeta_{jk^*} = 0.12$, and $\zeta_{jk^{**}} = 0.02$, meaning that the decision of setting $\hat{C}_{jk^{**}}$ to $0$ and $\hat{C}_{jk^*}$ to the posterior mean have low uncertainty (as $\zeta_{jk^{**}}=0.02$ and $\zeta_{jk^*}=0.12$).  Conversely, the decision of setting to $0$ the entry $\hat{C}_{jk}$ is very uncertain (since $\zeta_{jk}=0.98$).
\end{exa}

The element-wise PIP and the associated uncertainty index, $\zeta$, provide information on what coefficients are nonzero and quantify the uncertainty about this statement.
Instead, variable selection procedures are concerned with statistical techniques designed to identify and eliminate the subset of irrelevant covariates from the regression model.
The PIP-based approach previously described can easily address this issue in univariate settings, but multivariate regressions call for the adoption of different methods as a prediction can have different impacts on each response variable.
In fact, it may be possible that a covariate is irrelevant to predict a subset of the responses but exerts an influence on the remaining ones.
In the Supplement, we report a variable selection method based on a single scalar quantity analogous to the group lasso of \cite{Chakraborty2019} and another one based on the PIP previously defined.

To address the limitation of the element-wise approach, we propose a novel index that assesses the relative importance of each covariate and is computationally inexpensive, as it relies on the output of the SAVS computed at every iteration of the MCMC.

\begin{definition}[Relevance Index]
The relevance index of the $j$th covariate, $\RI_j$, with $j=1,\ldots,p$, is defined as:
\begin{equation}
\RI_j(k) = \frac{1}{M} \sum_{m=1}^M \I\big( N_{\bullet j}^{(m)}= kq \big),  \qquad  k=0,\frac{1}{q},\frac{2}{q},\ldots,1,   
\label{eq:definition_RI}
\end{equation}
where $N_{\bullet j}^{(m)} = \sum_{k=1}^q \I( \bar{C}_{jk}^{(m)} \neq 0)$ is the number of nonzero entries in the $j$th row of $\bar{C}$.
Notice that $\RI_j$ is a discrete distribution supported on $\mathcal{D} = \{ 0,1/q,2/q,\ldots,1 \}$, with mass at each $k\in\mathcal{D}$ given by Eq.~\eqref{eq:definition_RI}.
\end{definition}

The \RI\ can be interpreted as representing the distribution (across MCMC) of the ``relevance'' of a covariate, as measured by the share of response variables on which the covariate exerts a significant impact (i.e., nonzero). This motivates the support being the discrete grid between $0$ and $1$, with step size $1/q$.
Therefore, a strongly right-skewed distribution suggests that the covariate is irrelevant or relevant just to a small share of response variables, whereas a left-skewed distribution is typical of common predictors that impact all the responses.
An important feature of the \RI\ is that it easily allows for uncertainty quantification by means of the variance of the distribution.
For instance, take two right-skewed distributions, $\RI_j$ and $\RI_k$, characterised by different variances, $\sigma_j^2 > \sigma_k^2$. In this case, we have evidence in favour of considering both $x_j$ and $x_k$ irrelevant, but with higher uncertainty of this statement about $x_j$. Eventually, for sufficiently large variance, a binary statement about the irrelevance of the covariate may appear hazardous.

The approach to variable selection based on the \RI\ allows the assessment of the relevance of each variable and provides full uncertainty quantification.
In practice, one is often concerned with identifying and excluding irrelevant predictors. The shape and dispersion of the $\RI_j$ probability mass function intrinsically inform the impact exerted by the covariate in the overall model and the degree of uncertainty about this belief.  Besides, when the practitioner needs to make a binary decision about variable selection, it is possible to summarise the information content of the \RI\ to answer this question.
A possible heuristic for choosing whether to include or exclude a covariate relies on the tail distribution (or survival function) of $\RI_j$, as described in the following rule of thumb. 

\begin{definition}[Rule of thumb for variable selection]
Let $\overline{sr} \in [0,1]$ be the share of response variables on which the $j$th covariate is required to have a significant impact, and $\overline{p} \in (0,1)$ be the desired minimum probability.
Then, a \textit{rule of thumb} consists in excluding the $j$th covariate if the following statement is not satisfied:
\begin{equation}
S_{\RI_j}(\overline{sr}) = \mathbb{P}( \RI_j > \overline{sr} ) \geq \overline{p},
\label{eq:tail_RI}
\end{equation}
where the probability is computed with respect to the (discrete) distribution of $\RI_j$.  
\end{definition}

However, we remark that any measure for a binary decision will inevitably lose part of the information embedded in the \RI. Therefore, when variable selection is concerned, we suggest to \textit{jointly} interpret the output of the binary rule and the \RI\ distribution.
The following example illustrates the application of the rule of thumb to make a binary decision about variable selection.

\begin{figure}[t!h]
\centering
\renewcommand{\tabcolsep}{2pt}
\setlength{\abovecaptionskip}{1pt}
\begin{tabular}{c c c c c}
\multicolumn{2}{c}{\footnotesize{Variable 1}} & & \multicolumn{2}{c}{\footnotesize{Variable 2}} \\
\footnotesize{$\text{RI}_1$} & \footnotesize{$S_{\RI_1}(\cdot)$ }& & \footnotesize{$\text{RI}_2$ } & \footnotesize{$S_{\RI_2}(\cdot)$} \\
\includegraphics[trim=20 10 20 10,clip,scale=0.25]{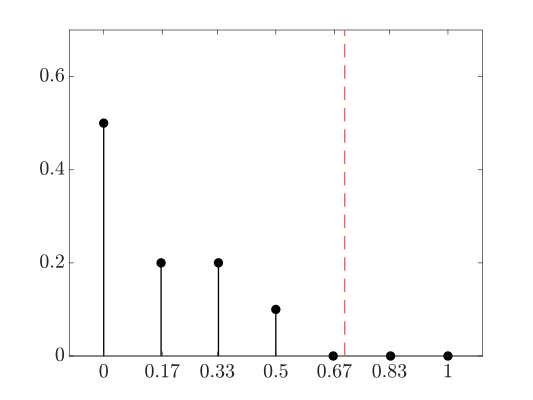} &
\includegraphics[trim=20 10 20 10,clip,scale=0.25]{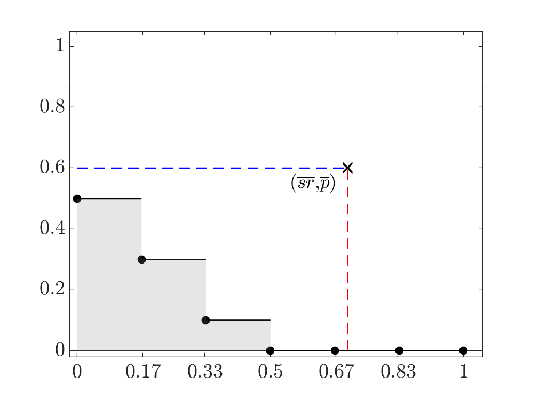} &
 \qquad &
\includegraphics[trim=20 10 20 10,clip,scale=0.25]{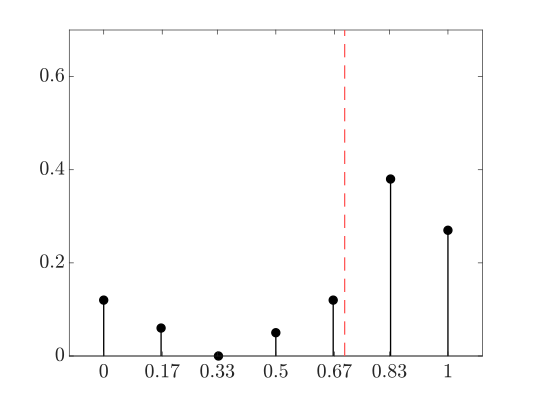} &
\includegraphics[trim=20 10 20 10,clip,scale=0.25]{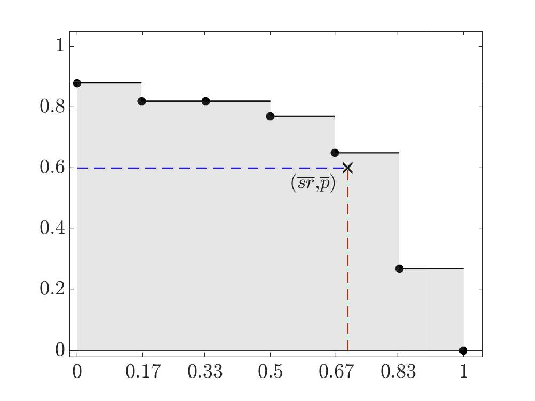}
\end{tabular}
\caption{Example of $\RI$ for two fictitious covariates: variable 1 (first and second plots) and variable 2 (third and fourth). The probability mass function (first and third plots) and the survival function (second and fourth) of the respective $\RI$ are reported for each covariate. Here we used $\overline{sr} = 0.70$ (red vertical line) and $\overline{p}=0.60$ (blue horizontal line). The area below the survival function is shaded: if the point $(\overline{sr},\overline{p})$ is located outside the shaded area, then the covariate is to be excluded.}
\label{fig:example_RI}
\end{figure}


\begin{exa}
We set $\overline{sr} = 0.70$ and $\overline{p} = 0.60$; thus, we require the \RI\ of the $j$th covariate to assign at least $0.60$ total probability mass on or above $0.70$.
In other words, this means that not to exclude the $j$th covariate, we require the probability of being relevant for more than the $70\%$ of responses to be at least $0.60$.
This example is represented in Figure~\ref{fig:example_RI}, which considers two covariates, characterised by a right-skewed and a left-skewed \RI. The associated survival functions are plotted together with the values of $\overline{sr}$ and $\overline{p}$ (vertical and horizontal dashed lines, respectively).
The \textit{rule of thumb} in Eq.~\eqref{eq:tail_RI} states that the covariate should be considered irrelevant if the point $(\overline{sr},\overline{p})$ lies above the curve of the survival function (i.e., outside the shaded area), and vice versa.
Therefore, in this simple example, the first covariate, which typically impacts a few response variables (as shown by the right-skewed \RI, first plot), is considered irrelevant (second plot). Conversely, the other covariate has a nonzero impact on most responses (see the left-skewed \RI\ in the third plot), and the irrelevance hypothesis is rejected for it (fourth plot).
However, notice that this decision does not account for the uncertainty quantified by the \RI. In this case, the variance of the \RI\ for the second covariate is high, thus implying that the selection decision for this variable should be taken with caution. Instead, the \RI\ for the first covariate is quite low, which suggests high confidence in the decision to exclude it from the model.
\end{exa}

\section{Simulation study} \label{sec:simulations}

We study the performance of the proposed reduced-rank model with the na\"ive (RRn) and the column-sharing (RRcs) priors across a range of simulation settings.  Our primary objectives in conducting this simulation study are twofold: firstly, to assess the efficacy of the model in accurately estimating the rank in varied settings, including different data generating processes (DGP), and secondly, to evaluate the recovery of the coefficient matrix under distinct scenarios.

The data was generated from the multivariate linear model $Y=XC_0 + E$, where we considered correlated and uncorrelated structures on the errors and regressors of the model. The rows of $X$ were independently drawn from $\mathcal{N}(0,\Sigma_X)$, with $\Sigma_X=I_p$ for independent regressors, and in the dependent case, the off-diagonal entries of $\Sigma_X$ are set to $0.5$.  The rows of $E$ were drawn from a zero mean multivariate normal distribution; under the assumption of uncorrelated errors, the covariance matrix is diagonal with elements sampled from $\mathcal{U}(0.5,1.75)$; for correlated errors, we consider the compound symmetry as in $X$. We work with centred responses and exclude the intercept term for simplicity.

Recalling the decomposition of the matrix of coefficients $C_0$ into the product of two matrices $A_0$ and $B_0$, the entries of both matrices are generated from the standard Gaussian, and their dimensions depend on the true rank $r_0 < \rmax$.  We consider two cases for the DGP of matrix $B_0$: non-sparse DGP, where the number of nonzero rows is equal to $p$, and sparse DGP, where the number of nonzero rows is $p^*<p$. Furthermore, our coefficient matrix estimation method is not limited to low-rank structures.  We have also tested our approach on an additional ``random zeros'' DGP, where a share of entries $z$ of the matrix $C$ is randomly set to zero (Figures~\ref{fig:n_z50_s},~\ref{fig:cs_z50_s}). The performance evaluation of the estimator $\hat{C}$ of the coefficient matrix was conducted by considering the mean squared error, \mbox{$\text{MSE}=\|\hat{C}-C_0\|_F^2/(pq)$}, where $\|\cdot\|_F$ is the Frobenius norm.\footnote{See section 3.2 of the Supplement for a CODA analysis for the posterior distribution of the rank and the estimates of the matrix $C$.}




\subsection{Simulation results}
\label{sec:sim_results}

The RRn tends to underestimate the value of the rank across the majority of settings, particularly 
when the number of covariates and responses increases. 
As more data become available, the posterior variance of the rank parameter shrinks, resulting in the concentration of the posterior distribution.  Therefore, the MCMC algorithm is further restricted in its ability to explore alternative values, resulting in a more pronounced underestimated rank.
However, in all these cases, the estimated coefficient matrix $\hat{C}$ is quite close to the true $C$ in all settings, succeeding in identifying the entries with high magnitude, and the estimates improve as the sample size grows.

\begin{figure}[t!]
    \centering
    \setlength{\abovecaptionskip}{2pt}
    \begin{subfigure}{0.32\linewidth}
    \setlength{\abovecaptionskip}{1.5pt}
        \includegraphics[trim=5mm 2mm 5mm 2mm,clip,width=\linewidth]{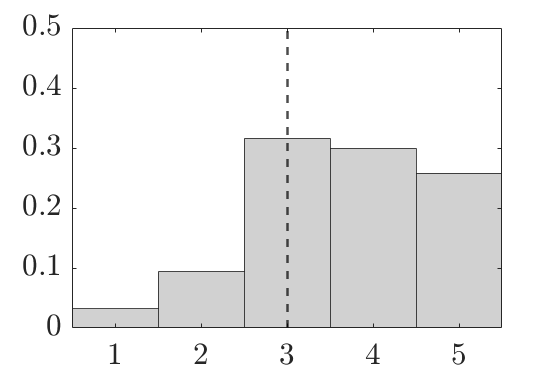}
        \caption{$n=50$}
    \end{subfigure}
    \begin{subfigure}{0.32\linewidth}
    \setlength{\abovecaptionskip}{1.5pt}
        \includegraphics[trim=5mm 2mm 5mm 2mm,clip,width=\linewidth]{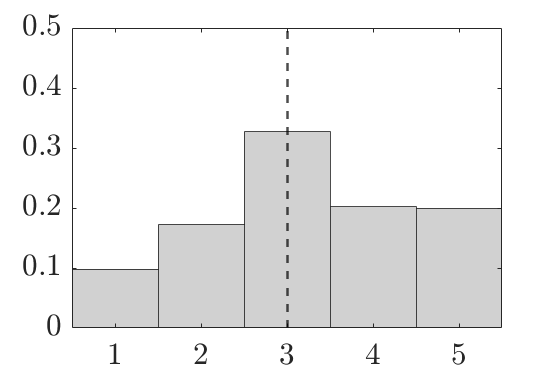}
        \caption{$n=100$}
    \end{subfigure}
    \begin{subfigure}{0.32\linewidth}
    \setlength{\abovecaptionskip}{1.5pt}
        \includegraphics[trim=5mm 2mm 5mm 2mm,clip,width=\linewidth]{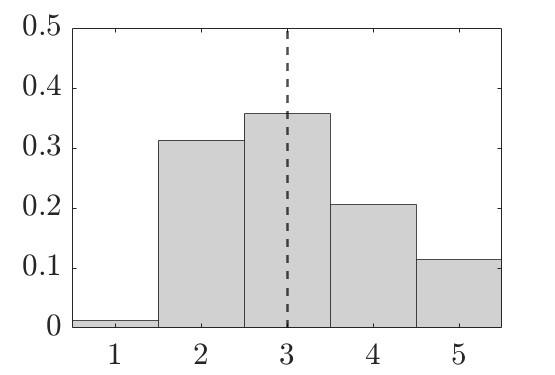}
        \caption{$n=500$}
    \end{subfigure}
    \caption{Posterior distribution of the rank in the non-sparse simulation setting with $(q,p)=(5,10)$ and true rank $r_0=3$ (dashed line) for different sample sizes, $n\in\{50,100,500\}$.}
    \label{fig:concentration}
\end{figure}

\begin{figure}[H]
  \centering
  \setlength{\abovecaptionskip}{2pt}
  \begin{subfigure}{0.45\linewidth}
    \centering
    \setlength{\abovecaptionskip}{1.5pt}
    \includegraphics[trim=34mm 5mm 17mm 0mm,clip,width=0.9\linewidth]{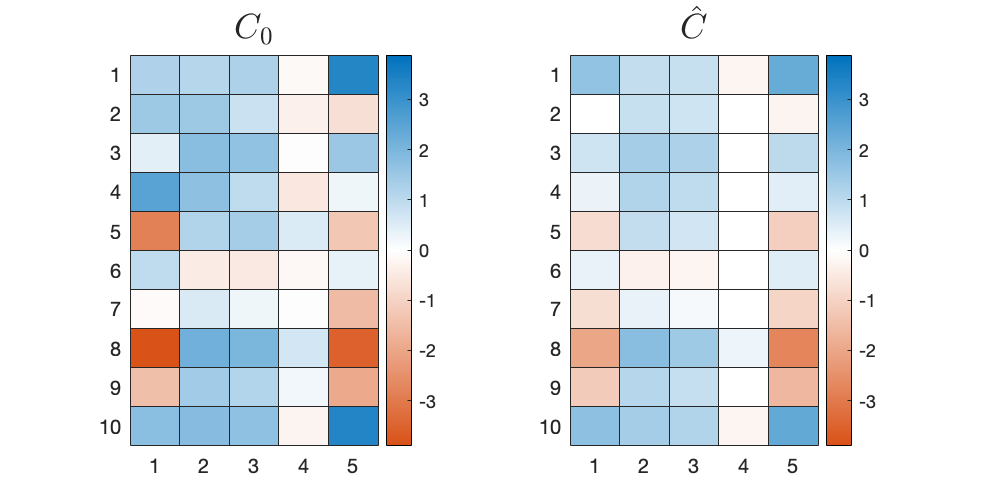}
    \caption{\footnotesize $\text{MSE}=0.097$}
    \label{fig:n_s10_s}
  \end{subfigure}
  \hspace{6pt}
  \begin{subfigure}{0.45\linewidth}
    \centering
    \setlength{\abovecaptionskip}{1.5pt}
    \includegraphics[trim=28mm 5mm 26mm 0mm,clip,width=0.9\linewidth]{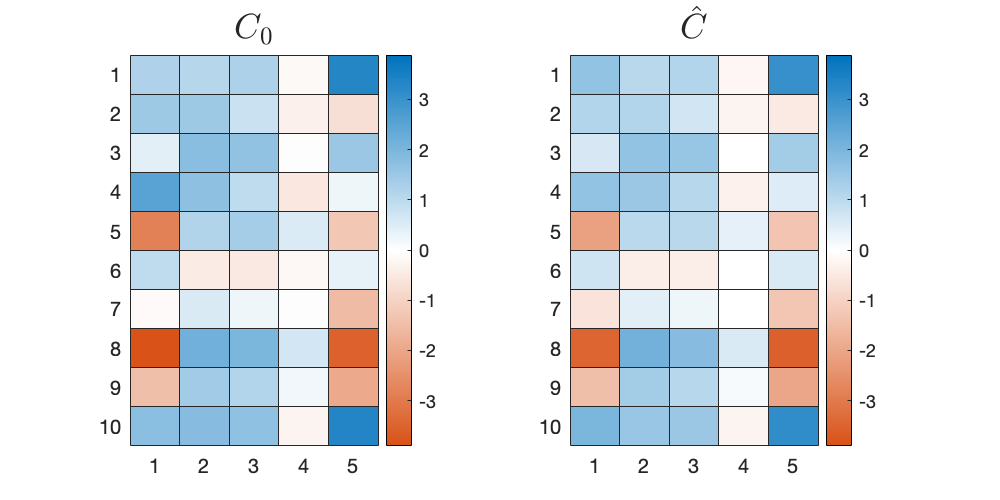}
    \caption{\footnotesize $\text{MSE}=0.036$}
    \label{fig:cs_s10_s}
  \end{subfigure}\\[-0.1cm]

  \begin{subfigure}{0.45\linewidth}
    \centering
    \setlength{\abovecaptionskip}{1.5pt}
    \includegraphics[trim=18mm 5mm 18mm 0mm,clip,width=0.9\linewidth]{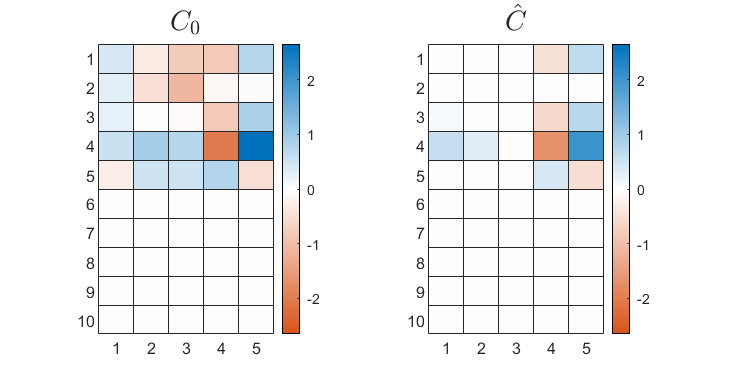}
    \caption{\footnotesize $\text{MSE}=0.044$}
    \label{fig:n_s5_s}
  \end{subfigure}
  \hspace{10pt}
  \begin{subfigure}{0.450\linewidth}
    \centering
    \setlength{\abovecaptionskip}{1.5pt}
    \includegraphics[trim=18mm 5mm 18mm 0mm,clip,width=0.9\linewidth]{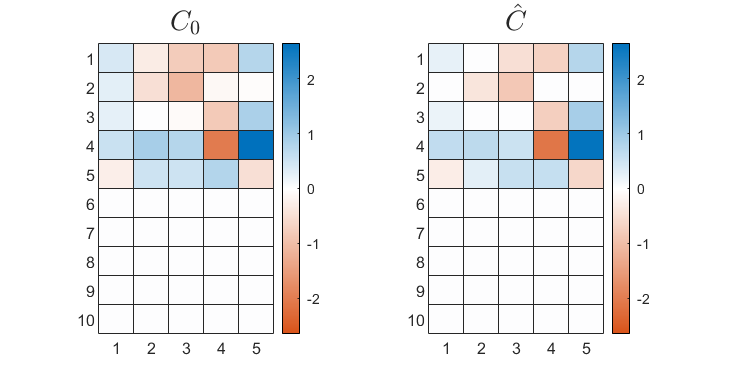}
    \caption{\footnotesize $\text{MSE}=0.016$}
    \label{fig:cs_s5_s}
  \end{subfigure}\\[-0.1cm]

  \begin{subfigure}{0.450\linewidth}
    \centering
    \setlength{\abovecaptionskip}{1.5pt}
    \includegraphics[trim=18mm 5mm 18mm 0mm,clip,width=0.9\linewidth]{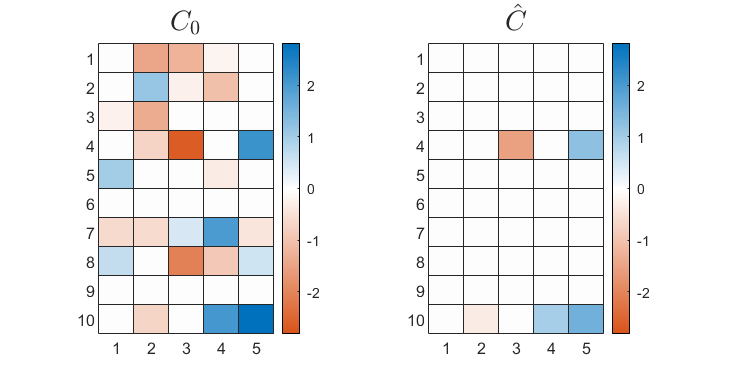}
    \caption{\footnotesize $\text{MSE}=0.100$}
    \label{fig:n_z50_s}
  \end{subfigure}
  \hspace{10pt}
  \begin{subfigure}{0.450\linewidth}
    \centering
    \setlength{\abovecaptionskip}{1.55pt}
    \includegraphics[trim=18mm 5mm 18mm 0mm,clip,width=0.9\linewidth]{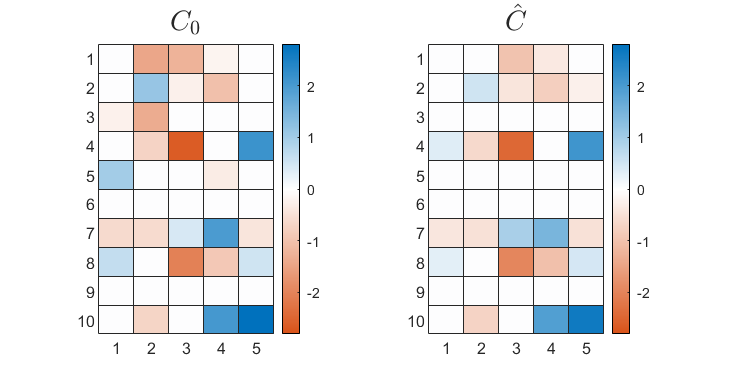}
    \caption{\footnotesize $\text{MSE}=0.051$}
    \label{fig:cs_z50_s}
  \end{subfigure}
  \caption{True ($C_0$) and estimated ($\hat{C}$) coefficient matrix. Data generated as described in Section~\ref{sec:simulations}, with $(q,p)=(5,10)$ and $n=100$; non-sparse matrix $B$ $(p^*=p)$ and $r_0=3$ (top); sparse matrix $B$ $(p^*=5)$ and $r_0=3$ (middle); $z=50\%$ of entries of $C$ set to $0$ (bottom).
  Results for the RRn (left - (a),(c),(e)) and RRcs (right - (b),(d)(f)), together with the MSE.} 
  \label{fig:C0vsChatsparse}
\end{figure}

After having shown the performance of the RRn, we provide evidence of the results for the RRcs approach.
The column-sharing exhibits superior performance to the na\"ive parametrization in terms of the MSE, convergence to the true rank and computational resources. 
The posterior distribution of $u$ tends to concentrates around the true value as $n$ increases (see Figure~\ref{fig:concentration}).  
However, when $B$ is sparse, the posterior distribution tends to put more mass on ranks smaller than $r_0$, resulting in a slight underestimation of the rank as the dimensionality of $(q,p)$ increases.  A possible motivation for these results is that having zero rows in $B$ implies sparsity in $C$ as well, which induces our method to prefer approximate $C$ with a small $r$ than to introduce additional parameters (higher $r$).  This feature of the model can be interpreted as favouring more parsimonious parametrizations.  Notice that the underestimation of $r$ in these cases has little impact, as $\hat{C}$ is nonetheless close to $C_0$ (Figure~\ref{fig:C0vsChatsparse}).  
RRcs consistently achieves a better mixing of the MCMC chain compared to RRn. Notably, in both parameterizations, sparsity contributes to the improved mixing.\footnote{Summary results comparing RRcs against RRn are included in Section 3.1 of the Supplement.}

The performance of RRcs deteriorates more rapidly for a fixed $n$ as the number of responses $q$ increases compared to the number of covariates $p$.  Conversely, the performance decays slower as $p$ increases and $q$ remains unchanged.  A change of $p$ to $p^\prime$ means $(p^\prime-p)\rmax$ more parameters to estimate.  However, when $q$ increases to $q^\prime$, the number of elements in $A$ and $\Sigma$ is directly affected, changing by $(q^\prime-q)\rmax$ and $(q^\prime-q)n$, respectively.  The crucial point is that $q$ represents the maximum rank in our context ($q\leq p)$.  Therefore, if $q$ increases, it increases the number of the mixture prior components, their respective weights, and the dimension of $B$.  Whether $B$ is sparse or not, it appears to have no impact on this trend.

%

We emphasise that our analysis relies on a sparse estimate of $C$, obtained using the SAVS method described in Section~\ref{sec:SAVS}. This estimation approach sets some entries to exact zeros, facilitating the variable selection.  This can be considered a binary classification problem, wherein the positive class encompasses the nonzero entries (representing significant coefficients). In contrast, the null entries (indicating irrelevant coefficients) belong to the negative category.  Consequently, our task involves identifying the position of zero and nonzero coefficients.  

To evaluate the performance of this classification task, we employ the Matthews correlation coefficient ($\MCC$), which is a more reliable statistical measure compared to commonly used metrics such as $\F_1$ score and accuracy \citep{chicco2020mccc}.  One notable advantage of $\MCC$, particularly relevant to our study, is its robustness in scenarios where one class contains significantly more samples than the other, thus addressing the issue of imbalanced datasets. Our synthetic data for $C_0$ repeatedly exhibits such imbalanced characteristics: in the non-sparse DGP when the majority of entries, if not all, deviate from zero, in the sparse DGP when only a few (or many) rows contain zero entries, and in the random zeros DGP when the percentage of zeros is low (or high).  This varying distribution of zeros in different scenarios allows us to assess the classification performance of our method under varying levels of sparsity and imbalance.

The classification model predicts the class for each data instance, assigning a predicted label (positive or negative) to each sample.  Depending on their actual class and their forecasted class, every sample is categorised into one of the following cases: true positives ($\TP$), true negatives ($\TN$), false positives ($\FP$) and false negatives ($\FN$). The $\MCC$ is given by:
\begin{equation}
    \MCC = \frac{\TP\times \TN-\FP\times\FN}{\sqrt{\left(\TP+\FP\right)\times\left(\TP+\FN\right)\times\left(\TN+\FP\right)\times\left(\TN+\FN\right)}} \in [-1,1].
    \label{eq:MCC}
\end{equation}
A value of $-1$ indicates poor performance, while a value of $1$ represents the highest level of performance. $\MCC$ produces a high-quality score only if the prediction obtained good results in all categories $(\TP,\TN,\FP,\FN)$, proportionally both to the size of positive and negative elements in the dataset.
Instead, the measure is undefined when any of the factors in the denominator is $0$, and specific mathematical reasoning should be considered.  For instance, if $C_0$ comprises only nonzero (zero) entries, and they are all correctly identified in $\hat{C}$, then $\TN=0$ ($\TP=0$) and $\FN=0$ ($\FP=0$), resulting in an undefined $\MCC$. Nonetheless, the classifier successfully identifies all samples in this case and achieves a perfect score of $1$. If all samples belong to the same class and are all incorrectly predicted, then $\MCC=-1$. We are left to study the cases when mixed samples are categorised into the same class or homogeneous samples are allocated to mixed classes. In either case, the correlation coefficient can be approximated by $0$ \citep[see][for a detailed description]{chicco2020mccc}.

We report the $\MCC$, the true positive rate (TPR = \TP/(\TP+\FN)) of correctly identified nonzero entries, and the false negative rate (FNR = \FN/(\TP+\FN)) of incorrectly identified zero entries across different simulation settings in Table~\ref{tab:MCC}. The non-negative $\MCC$s obtained by our methods suggest a good performance in the estimation of the coefficient matrix, favouring once more the RRcs parametrization over RRn. Having attained $\MCC$s close to $1$, we show the robustness of the former method in accurately approximating the $C$ matrix while incorporating sparsity in its estimation. The standard DGP illustrates the need for the corrections to the formula of the $\MCC$ when undefined since $C_0$ consisted of only nonzero entries; meanwhile, the estimated $\hat{C}$ had mixed values.  The algorithm established $16\%$ of the entries as $0$, thus allowing for a sparse model with enhanced interpretability.

\begin{table}[t!]
\centering
\resizebox{0.8\linewidth}{!}{%
\begin{tabular}{@{}cc|ccc|ccc@{}}
\toprule
\multicolumn{2}{c|}{\multirow{2}{*}{\textbf{DGP}}} & \multicolumn{3}{c|}{\textbf{RRn}} & \multicolumn{3}{c}{\textbf{RRcs}} \\
\multicolumn{2}{c|}{} & \textbf{MCC} & \textbf{TPR} & \textbf{FNR} & \textbf{MCC} & \textbf{TPR} & \textbf{FNR} \\ \midrule
\textbf{Standard} & $p^*=10$ & --- & 0.84 & 0.16 & --- & 0.94 & 0.06 \\ \midrule
\multirow{4}{*}{\textbf{Sparse}} & $p^*=2$ & 0.00 & 0.00 & 1.00 & 0.81 & 0.70 & 0.30 \\
 & $p^*=5$ & 0.53 & 0.44 & 0.56 & 0.78 & 0.76 & 0.24 \\
 & $p^*=8$ & 0.13 & 0.08 & 0.92 & 0.54 & 0.68 & 0.32 \\
 & $p^*=9$ & 0.38 & 0.62 & 0.38 & 0.83 & 0.96 & 0.04 \\ \midrule
\multirow{4}{*}{\textbf{Random zeros}} & $z=0.20$ & 0.13 & 0.08 & 0.92 & 0.73 & 0.85 & 0.15 \\
 & $z=0.50$ & 0.33 & 0.20 & 0.80 & 0.73 & 0.80 & 0.20 \\
 & $z=0.80$ & 0.00 & 0.00 & 1.00 & 0.74 & 0.60 & 0.40 \\
 & $z=0.90$ & 0.00 & 0.00 & 1.00 & 0.88 & 0.80 & 0.20 \\ \bottomrule
\end{tabular}%
}
\caption{Measures of association between the true coefficient matrix $C_0$ and the sparse estimate $\hat{C}$ in the setting $(q,p)=(5,10)$, $n=100$, $r_0=3$ in the standard and sparse scenarios.  $p^*$ represents the number of nonzero rows in $B$ (and consequently in $C_0$), and $z$ is the proportion of randomly allocated zero entries in $C_0$.  For RRn and RRcs, we report the MCC, TPR and FNR.
}
\label{tab:MCC}
\end{table}

\subsection{Comparison to other methods}
\label{sec:sim_comparison}

The performance of the proposed mixture prior RRcs is comparable to other state-of-the-art methodologies in reduced-rank regression, as we showcase in this section. We evaluate our approach against the frequentist methods of \cite{Chen2013} and \cite{she2017robust}. The former utilises an adaptive nuclear norm penalisation approach (ANN), where the rank is estimated by the threshold of singular values. The latter proposes a robust reduced-rank regression approach (RRRR), which requires the user to choose the optimal rank, a task that is achieved through a suggested criterion.

The data was generated as outlined in Section~\ref{sec:simulations}. 
For each configuration, we conducted 20 replications of the experiment. Consequently, the results presented reflect the average estimated rank and MSE across these repetitions.
Notably, our method achieves similar outcomes as both ANN and RRRR, particularly outperforming them when the sample size is $n=50$ and the dimensions of $q$ and $p$ increase (see Table~\ref{tab:comp_n50_sparse} and Table~\ref{tab:comp_n50_nonsparse}). We defer additional results to the Supplement.

\begin{table}[h!t]
\centering
\resizebox{0.82\textwidth}{!}{%
\begin{tabular}{@{}cccc|ccc|ccc@{}}
\toprule
                           &  &                             &           & \multicolumn{3}{c|}{$\Sigma_{ind}$} & \multicolumn{3}{c}{$\Sigma_{corr}$} \\
$\mathbf{(q,p)}$ &
  $\mathbf{r_0}$ &
  $\mathbf{X}$ &
  \textbf{Measure} &
  \textbf{ANN} &
  \textbf{RRRR} &
  \textbf{RRcs} &
  \textbf{ANN} &
  \textbf{RRRR} &
  \textbf{RRcs} \\ \midrule[0.1em]
\multirow{8}{*}{$(5,15)$} &
  \multirow{4}{*}{3} &
  \multirow{2}{*}{$X_{ind}$} &
  $\hat{r}$ &
  2.25 &
  2.4 &
  3.15 &
  2.25 &
  2.25 &
  2.8 \\
                           &                       &                             & MSE       & 0.028     & 0.026        & 0.019    & 0.030      & 0.030       & 0.022    \\ \cmidrule(l){3-10} 
                           &                       & \multirow{2}{*}{$X_{corr}$} & $\hat{r}$ & 2.20       & 2.15         & 3.05     & 2.05       & 1.85        & 2.45     \\
                           &                       &                             & MSE       & 0.046     & 0.051        & 0.036    & 0.042      & 0.049       & 0.031    \\ \cmidrule(l){2-10} 
                           & \multirow{4}{*}{5}    & \multirow{2}{*}{$X_{ind}$}  & $\hat{r}$ & 3.10       & 3.25         & 4.45     & 3.05       & 3.10         & 3.75     \\
                           &                       &                             & MSE       & 0.038     & 0.034        & 0.017    & 0.034      & 0.036       & 0.022    \\ \cmidrule(l){3-10} 
                           &                       & \multirow{2}{*}{$X_{corr}$} & $\hat{r}$ & 2.60       & 2.90          & 4.50      & 3.05       & 3.00           & 3.80      \\
                           &                       &                             & MSE       & 0.071     & 0.062        & 0.035    & 0.059      & 0.064       & 0.050    \\ \midrule
\multirow{8}{*}{$(5,50)$}  & \multirow{4}{*}{3}    & \multirow{2}{*}{$X_{ind}$}  & $\hat{r}$ & 0.05      & 5.00            & 2.35     & 0.05       & 5.00           & 1.65     \\
                           &                       &                             & MSE       & 1.754     & 16.800       & 0.108    & 0.451      & 4.279       & 0.083    \\ \cmidrule(l){3-10} 
                           &                       & \multirow{2}{*}{$X_{corr}$} & $\hat{r}$ & 0.25      & 5.00            & 2.80      & 0.20        & 5.00           & 1.9      \\
                           &                       &                             & MSE       & 0.984     & 34.732       & 0.068    & 0.954      & 37.709      & 0.100    \\ \cmidrule(l){2-10} 
                           & \multirow{4}{*}{5}    & \multirow{2}{*}{$X_{ind}$}  & $\hat{r}$ & 0.10       & 5.00            & 4.40      & 0.00          & 5.00           & 3.60      \\
                           &                       &                             & MSE       & 3.178     & 1843.949     & 0.043    & 0.930      & 25.676      & 0.058    \\ \cmidrule(l){3-10} 
                           &                       & \multirow{2}{*}{$X_{corr}$} & $\hat{r}$ & 0.25      & 5.00            & 4.90      & 0.00          & 5.00           & 4.35     \\
                           &                       &                             & MSE       & 0.859     & 19.219       & 0.042    & 1.056      & 39.181      & 0.056    \\ \midrule
\multirow{8}{*}{$(10,50)$} & \multirow{4}{*}{3}    & \multirow{2}{*}{$X_{ind}$}  & $\hat{r}$ & 1.65      & 10.00           & 1.00        & 1.65       & 10.00          & 1.1      \\
                           &                       &                             & MSE       & 5.800     & 165.912      & 0.243    & 23.358     & 415.744     & 0.226    \\ \cmidrule(l){3-10} 
                           &                       & \multirow{2}{*}{$X_{corr}$} & $\hat{r}$ & 0.95      & 10.00           & 1.20      & 1.65       & 10.00          & 1.05     \\
                           &                       &                             & MSE       & 2.824     & 15.210       & 0.212    & 2.787      & 49.317      & 0.289    \\ \cmidrule(l){2-10} 
                           & \multirow{4}{*}{5}    & \multirow{2}{*}{$X_{ind}$}  & $\hat{r}$ & 0.15      & 10.00           & 1.10      & 0.05       & 10.00          & 1.00        \\
                           &                       &                             & MSE       & 1.034     & 100.673      & 0.531    & 0.891      & 18.591      & 0.503    \\ \cmidrule(l){3-10} 
                           &                       & \multirow{2}{*}{$X_{corr}$} & $\hat{r}$ & 0.40       & 10.00           & 1.35     & 0.45       & 10.00          & 1.05     \\
                           &                       &                             & MSE       & 0.990     & 78.371       & 0.538    & 3.082      & 52.587      & 0.543   \\
                           \bottomrule
\end{tabular}%
}
\caption{Comparison of the estimated rank ($\hat{r}$) and mean squared error (MSE) obtained by RRcs against ANN \citep{Chen2013} and RRRR \citep{she2017robust} for different values of $(q,p)$ and true rank $r_0$. In all settings, $n=50$, and the DGP is sparse with $p^*=5$ if $p=15$, while $p^*=10$ if $p=50$. We present the average estimates over 20 repetitions for independent errors ($\Sigma_{ind}$), correlated errors ($\Sigma_{corr}$), independent regressors ($X_{ind}$), and correlated regressors ($X_{corr}$).}
\label{tab:comp_n50_sparse}
\end{table}

\begin{table}[h!t]
\centering
\resizebox{0.82\textwidth}{!}{%
\begin{tabular}{@{}cccc|ccc|ccc@{}}
\toprule
 &
   &
   &
   &
  \multicolumn{3}{c|}{$\Sigma_{ind}$} &
  \multicolumn{3}{c}{$\Sigma_{corr}$} \\ 
\textbf{$\mathbf{(q,p)}$} &
  \textbf{$\mathbf{r_0}$} &
  \textbf{$\mathbf{X}$} &
  \multicolumn{1}{c|}{\textbf{Measure}} &
  \textbf{ANN} &
  \textbf{RRRR} &
  \multicolumn{1}{c|}{\textbf{RRcs}} &
  \textbf{ANN} &
  \textbf{RRRR} &
  \textbf{RRcs} \\ \midrule[0.1em]
\multirow{8}{*}{$(5,15)$} &
  \multirow{4}{*}{3} &
  \multirow{2}{*}{$X_{ind}$} &
  \multicolumn{1}{c|}{$\hat{r}$} &
  2.90 &
  2.80 &
  \multicolumn{1}{c|}{3.65} &
  2.95 &
  3.00 &
  3.35 \\
 &
   &
   &
  \multicolumn{1}{c|}{MSE} &
  0.022 &
  0.026 &
  \multicolumn{1}{c|}{0.129} &
  0.023 &
  0.023 &
  0.089 \\ \cmidrule(l){3-10} 
 &
   &
  \multirow{2}{*}{$X_{corr}$} &
  \multicolumn{1}{c|}{$\hat{r}$} &
  2.90 &
  2.90 &
  \multicolumn{1}{c|}{3.9} &
  3.00 &
  3.00 &
  3.05 \\
 &
   &
   &
  \multicolumn{1}{c|}{MSE} &
  0.040 &
  0.042 &
  \multicolumn{1}{c|}{0.102} &
  0.037 &
  0.041 &
  0.195 \\ \cmidrule(l){2-10} 
 &
  \multirow{4}{*}{5} &
  \multirow{2}{*}{$X_{ind}$} &
  \multicolumn{1}{c|}{$\hat{r}$} &
  4.00 &
  4.40 &
  \multicolumn{1}{c|}{4.85} &
  3.95 &
  4.35 &
  4.90 \\
 &
   &
   &
  \multicolumn{1}{c|}{MSE} &
  0.059 &
  0.034 &
  \multicolumn{1}{c|}{0.151} &
  0.044 &
  0.031 &
  0.102 \\ \cmidrule(l){3-10} 
 &
   &
  \multirow{2}{*}{$X_{corr}$} &
  \multicolumn{1}{c|}{$\hat{r}$} &
  3.90 &
  4.40 &
  \multicolumn{1}{c|}{4.95} &
  4.00 &
  4.35 &
  5.00 \\
 &
   &
   &
  \multicolumn{1}{c|}{MSE} &
  0.092 &
  0.066 &
  \multicolumn{1}{c|}{0.174} &
  0.074 &
  0.059 &
  0.128 \\ \midrule
\multirow{8}{*}{$(5,50)$} &
  \multirow{4}{*}{3} &
  \multirow{2}{*}{$X_{ind}$} &
  \multicolumn{1}{c|}{$\hat{r}$} &
  0.65 &
  5.00 &
  \multicolumn{1}{c|}{1.7} &
  1.55 &
  5.00 &
  1.40 \\
 &
   &
   &
  \multicolumn{1}{c|}{MSE} &
  6.912 &
  20.468 &
  \multicolumn{1}{c|}{1.034} &
  207.621 &
  15.594 &
  1.292 \\ \cmidrule(l){3-10} 
 &
   &
  \multirow{2}{*}{$X_{corr}$} &
  \multicolumn{1}{c|}{$\hat{r}$} &
  1.00 &
  5.00 &
  \multicolumn{1}{c|}{3.15} &
  1.40 &
  5.00 &
  2.00 \\
 &
   &
   &
  \multicolumn{1}{c|}{MSE} &
  19.369 &
  41.381 &
  \multicolumn{1}{c|}{1.050} &
  6.361 &
  13.865 &
  1.164 \\ \cmidrule(l){2-10} 
 &
  \multirow{4}{*}{5} &
  \multirow{2}{*}{$X_{ind}$} &
  \multicolumn{1}{c|}{$\hat{r}$} &
  0.00 &
  5.00 &
  \multicolumn{1}{c|}{5} &
  0.20 &
  5.00 &
  4.80 \\
 &
   &
   &
  \multicolumn{1}{c|}{MSE} &
  5.079 &
  32.540 &
  \multicolumn{1}{c|}{1.456} &
  4.656 &
  1399.716 &
  1.653 \\ \cmidrule(l){3-10} 
 &
   &
  \multirow{2}{*}{$X_{corr}$} &
  \multicolumn{1}{c|}{$\hat{r}$} &
  0.15 &
  5.00 &
  \multicolumn{1}{c|}{5} &
  0.20 &
  5.00 &
  5.00 \\
 &
   &
   &
  \multicolumn{1}{c|}{MSE} &
  22.558 &
  42.882 &
  \multicolumn{1}{c|}{1.869} &
  5.952 &
  176.258 &
  1.874 \\ \midrule
\multirow{8}{*}{$(10,50)$} &
  \multirow{4}{*}{3} &
  \multirow{2}{*}{$X_{ind}$} &
  \multicolumn{1}{c|}{$\hat{r}$} &
  3.00 &
  10.00 &
  \multicolumn{1}{c|}{1.65} &
  3.15 &
  10.00 &
  1.55 \\
 &
   &
   &
  \multicolumn{1}{c|}{MSE} &
  3473.521 &
  1208.176 &
  \multicolumn{1}{c|}{1.011} &
  58.882 &
  215.646 &
  1.399 \\ \cmidrule(l){3-10} 
 &
   &
  \multirow{2}{*}{$X_{corr}$} &
  \multicolumn{1}{c|}{$\hat{r}$} &
  3.00 &
  10.00 &
  \multicolumn{1}{c|}{1.6} &
  3.20 &
  10.00 &
  1.45 \\
 &
   &
   &
  \multicolumn{1}{c|}{MSE} &
  639.223 &
  83.254 &
  \multicolumn{1}{c|}{1.394} &
  6.677 &
  95.078 &
  1.544 \\ \cmidrule(l){2-10} 
 &
  \multirow{4}{*}{5} &
  \multirow{2}{*}{$X_{ind}$} &
  \multicolumn{1}{c|}{$\hat{r}$} &
  2.50 &
  10.00 &
  \multicolumn{1}{c|}{1.7} &
  3.65 &
  10.00 &
  1.15 \\
 &
   &
   &
  \multicolumn{1}{c|}{MSE} &
  53.516 &
  319.769 &
  \multicolumn{1}{c|}{3.203} &
  12.321 &
  31.421 &
  3.859 \\ \cmidrule(l){3-10} 
 &
   &
  \multirow{2}{*}{$X_{corr}$} &
  \multicolumn{1}{c|}{$\hat{r}$} &
  1.95 &
  10.00 &
  \multicolumn{1}{c|}{1.5} &
  3.00 &
  10.00 &
  1.55 \\
 &
   &
   &
  \multicolumn{1}{c|}{MSE} &
  12.253 &
  50.605 &
  \multicolumn{1}{c|}{3.370} &
  12.697 &
  93.531 &
  3.777 \\ \bottomrule
\end{tabular}%
}
\caption{Comparison of the estimated rank ($\hat{r}$) and mean squared error (MSE) obtained by RRcs against ANN \citep{Chen2013} and RRRR \citep{she2017robust} for different values of $(q,p)$ and true rank $r_0$. In all settings, $n=50$, and the DGP is non-sparse with $p^*=p$. We present the average estimates over 20 repetitions for independent errors ($\Sigma_{ind}$), correlated errors ($\Sigma_{corr}$), independent regressors ($X_{ind}$), and correlated regressors ($X_{corr}$).}
\label{tab:comp_n50_nonsparse}
\end{table}

\section{Applications} \label{sec:application}

This section showcases the efficacy of our proposed method through its practical implementation on two actual datasets.  By providing empirical demonstrations of the method, we illustrate its value and effectiveness in addressing real-life situations.

\subsection{Chemical composition of tobacco}

The dataset on the chemical composition of $n=25$ tobacco leaf samples, taken from \cite{anderson1952statistical}, illustrates how our proposed methodology produces comparable results as those obtained in the literature with further advantages addressed subsequently.  

Tobacco leaves are made up of organic and inorganic chemical constituents, and the typical interest is investigating the relationship between certain constituents.  The $p=6$ covariates are per cent nitrogen, per cent chlorine, per cent potassium, per cent phosphorus, per cent calcium, and per cent magnesium.  We also include an intercept term. The $q=3$ response variables are the rate of cigarette burn in inches per 1,000 seconds, the percentage of sugar in the leaf, and the percentage of nicotine in the leaf.  

\begin{figure}[bt]
    \centering
    \setlength{\abovecaptionskip}{2pt}
    \begin{subfigure}{0.35\linewidth}
        \setlength{\abovecaptionskip}{2pt}
        \includegraphics[width=\linewidth]{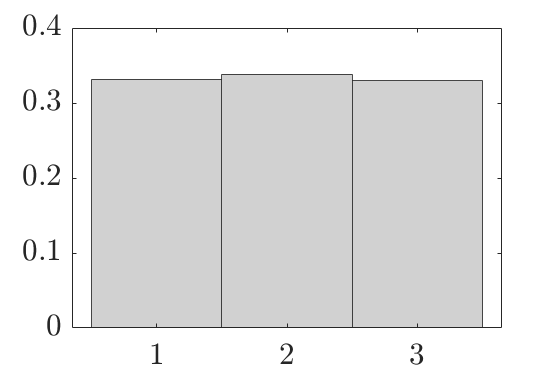}
        \caption{Posterior distribution of $u$.}
        \label{fig:tobacco_u}
    \end{subfigure}
    \begin{subfigure}{0.57\linewidth}
        \setlength{\abovecaptionskip}{2pt}
        \includegraphics[width=\linewidth]{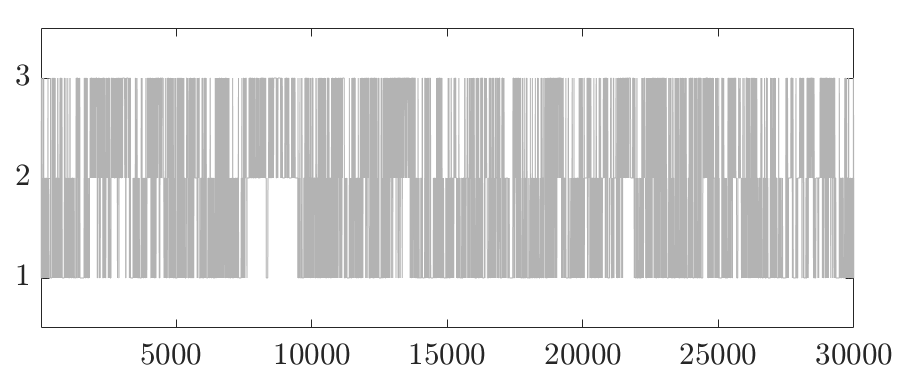}
        \caption{MCMC chain of $u$.}
        \label{fig:tobacco_chain}
    \end{subfigure}
    \caption{Tobacco dataset: posterior distribution (panel a) and MCMC chain after burn-in (panel b) of the rank $u$ for the coefficient matrix $C$.}
    \label{fig:tobacco}
\end{figure}

Under the column-sharing parametrization, the posterior distribution for the rank, as illustrated in Figure~\ref{fig:tobacco}, shows no significant difference between any of the three values, identifying a potential uniform posterior distribution for $r$.  Our result agrees with the findings of \citet{Izenman2008book}, who uses the rank trace method.  This procedure first estimates the coefficient matrix for each rank that minimises a weighted sum of squares criterion and the residual covariance matrix.  Then, the rank is gradually increased, and the entries in both matrices will change significantly until the true rank is reached, where the matrices will stabilise.  The change in the residual covariance matrix at each increment of $r$ is plotted against the change in $C$ in a scatterplot.  Finally, the rank of $C$ is assessed as the smallest rank for which the differences are close to $0$.  The rank-trace plot applied to the tobacco dataset shows that the rank-$1$ or rank-$2$ solutions have no discernible difference between them and the full-rank solution.
The main drawback of this method is that conclusions on the effective dimensionality of the multivariate regression involve subjective judgement and visual interpretation of the rank trace plot \citep[see][for more details]{Izenman2008book}.

To statistically validate our hypothesis that the posterior distribution of the rank is uniform, we conduct a Pearson's chi-squared test for goodness of fit, obtaining a $p$-value of $0.1595$, thus implying the non-rejection of the null hypothesis.  Formal statistical testing for uniformity yields a more objective result as opposed to relying purely on visual analysis.  Moreover, our method requires the estimation of one model, whereas the rank trace performs the estimation of three distinct models.  Besides the estimates of the rank and the coefficient matrix, we provide in conjunction uncertainty quantification as an objective means of analysis, moving away from reliance on subjective judgement.

Regarding the estimated coefficient matrix for each rank, our algorithm produces consistent estimates compared to those reported in \citet{Izenman2008book}, with the aforementioned advantages.

\subsection{COMBO-17 galaxy photometric data}

The second dataset consists of a subset of a public catalogue of astronomical objects, COMBO-17 (Classifying Objects by Medium-Band Observations in $17$ filters), a project of international collaboration aimed at exploring the evolution of galaxies \citep{Wolf2004galaxy}.  The present dataset is utilised herein to illustrate the proposed variable selection procedure and the associated uncertainty.\footnote{An additional forecasting exercise is presented in Section 4.3 of the Supplement.}  The methodology serves as a reliable means of informing decision-making regarding selecting covariates, offering both suggestions and quantifying uncertainty in such selections.

The original dataset consists of $63,501$ objects in the area of the sky named Chandra Deep Field South with brightness measurements in $17$ passbands from $350$ to $930$ nm.  We restrict the analysis to $3,438$ objects, all classified as ``Galaxies'' by \citet{Wolf2004galaxy}, and with no missing values for any of the $65$ variables.  The measurement errors and five redundant variables were omitted, resulting in a total of 29 variables divided into $p=23$ covariates and $q=6$ responses, as done in \citet{Izenman2008book}.  Regarding the covariates, $10$ variables correspond to the absolute magnitudes of the galaxy in $10$ bands, while the remaining variables are the observed brightness in $13$ bands across the range $420-915$ nm.  Meanwhile, the responses are the total R-band magnitude, the aperture difference of the R-band, the central surface brightness in the R-band, two redshift estimates, and the reduced chi-squared value of the best-fitting template galaxy spectrum.

Given that the whole subset of galaxies consists of a considerable number of $n = 3,438$ observations, the level of estimation uncertainty is likely to be quite small.
Therefore, motivated by the intention of highlighting the ability to quantify the uncertainty of the proposed BRECS method, we apply it to a sub-sample of $n=500$ observations randomly selected from the entire subset of data.
The rank selection, sparse estimation, and variable selection procedures are also repeated for the full sample and for other randomly chosen sub-samples of size $n=500$ and $n=1,500$ (see the Supplement). In line with expectations, we find that the use of larger samples reduces the uncertainty, but the key insights about the advantages of using the BRECS method are unaltered.

\begin{figure}[t!]
    \centering
    \setlength{\abovecaptionskip}{2pt}
    \begin{subfigure}{0.32\linewidth}
        \setlength{\abovecaptionskip}{1.5pt}
        \includegraphics[trim=8mm 10mm 3mm 12mm,clip,width=0.9\linewidth]{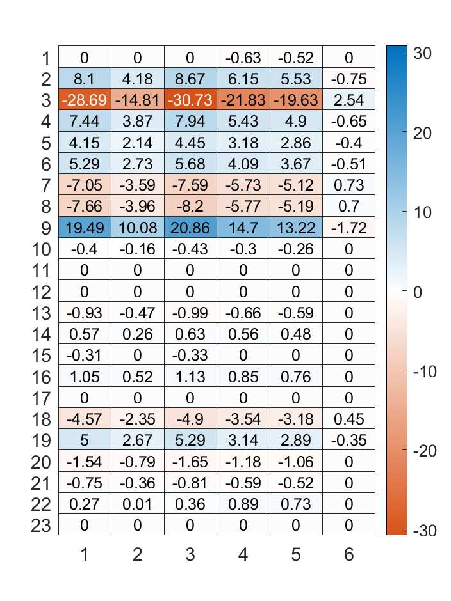}
        \caption{Sparse estimate $\hat{C}$.}
        \label{fig:galaxy_Chat}
    \end{subfigure}
    \hfill
    \begin{subfigure}{0.32\linewidth}
        \setlength{\abovecaptionskip}{1.5pt}
        \includegraphics[width=0.9\linewidth]{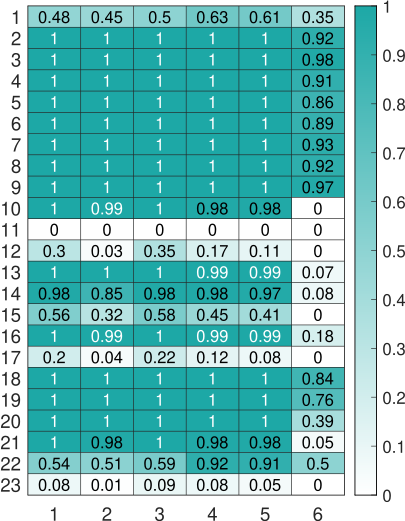}
        \caption{PIP of $C_{jk}$.}
        \label{fig:galaxy_PIP}
    \end{subfigure}
    \hfill
    \begin{subfigure}{0.32\linewidth}
        \setlength{\abovecaptionskip}{1.5pt}
        \includegraphics[width=0.9\linewidth]{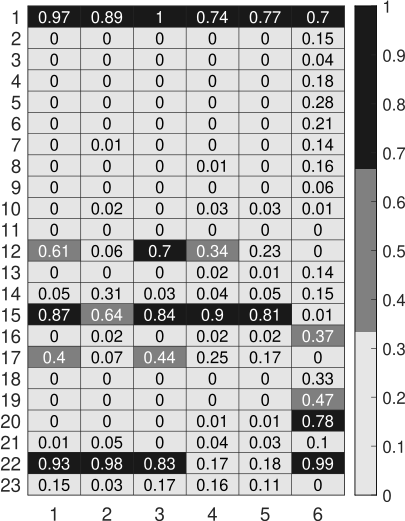}
        \caption{PIP uncertainty index $\zeta_{jk}$.}
        \label{fig:galaxy_zeta}
    \end{subfigure}
    \caption{Sparse estimate $\hat{C}$ of the coefficient matrix $C$ of the linear regression model with the galaxy dataset of $n=500$ observations (panel a), the uncertainty about this estimation through the posterior inclusion probabilities (panel b), and the PIP uncertainty index, in a grey-colour scale according to low ($\zeta_{jk} \leq 1/3$), medium ($1/3 < \zeta_{jk} \leq 2/3$), or high ($\zeta_{jk} > 2/3$) uncertainty (panel c).}
    \label{fig:galaxy_C}
\end{figure}

The posterior distribution of the rank is right-skewed and achieves the maximum (MAP) at $\hat{u} = 2$. Considering all the $3,438$ observations, the posterior distribution is more concentrated around the same maximum point (see the Supplement for further details).


The effect of the covariates on the responses emphasises the importance of estimating the matrix $C$.  For any pair $(jk)$, a zero entry $\hat{C}_{jk}=0$ means that there is no association between the $j$th covariate and the $k$th response.  Therefore, zero entries facilitate interpretation as the nonzero rows of $C$ identify the covariates that influence at least one response.  We obtain a sparse estimate of $C_{jk}$ by applying the SAVS algorithm at each iteration of the MCMC, computing its posterior inclusion probability $\PIP_{jk}$, and set the element to $0$ if $\PIP_{jk}\leq 0.5$ or to its posterior mean otherwise, as described in Section~\ref{sec:posterior}.  The matrix of PIPs takes values in $\left[0,1\right]$, and the elements of $C$ with equal or less than $50\%$ probability of inclusion are set to $0$.  We emphasise that even though some entries of the sparse estimate $\hat{C}$ are $0$, the probability of including them is not exactly $0$.  At first inspection, the $12$th covariate, the observed brightness of the galaxy in the corresponding band, is to be ruled out of the model since its coefficients are only zeros (see Figure~\ref{fig:galaxy_Chat}).  However, not all of the PIPs of covariate $12$ are close to $0$, especially $\PIP_{12,1}=0.30$ and $\PIP_{12,3}=0.35$ (see Figure~\ref{fig:galaxy_PIP}).  The estimated coefficient matrix $\hat{C}$ of the galaxy dataset exposes $4$ complete zero rows, thus reducing the number of covariates exerting a significant impact to $19$.

\begin{figure}[t!]
    \centering
    \setlength{\abovecaptionskip}{2pt}
    \begin{subfigure}{0.32\linewidth}
        \includegraphics[trim=6mm 2mm 7mm 2mm,clip,width=\linewidth]{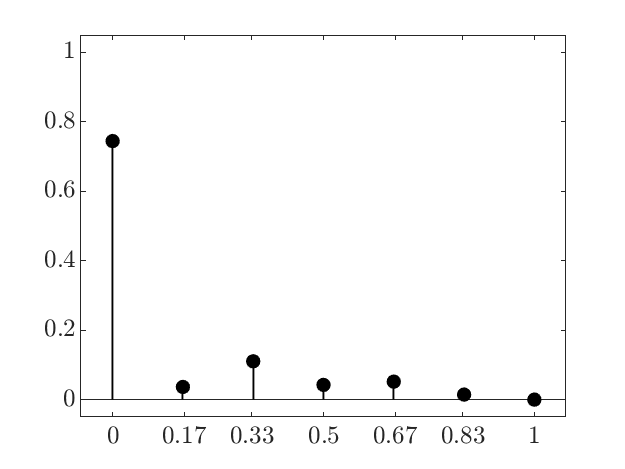}
    \end{subfigure}
    \begin{subfigure}{0.32\linewidth}
        \includegraphics[trim=6mm 2mm 7mm 2mm,clip,width=\linewidth]{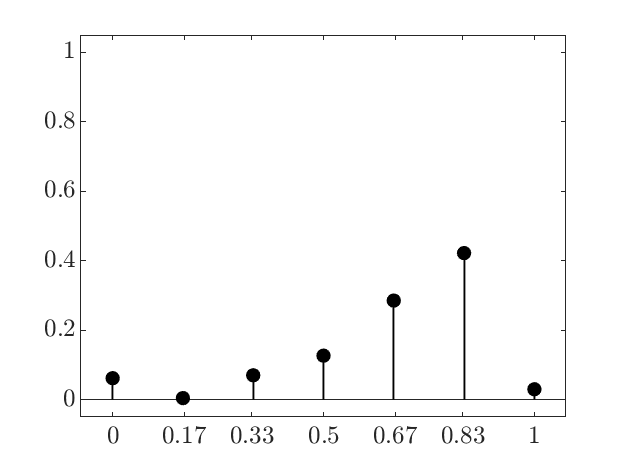}
    \end{subfigure}
    \begin{subfigure}{0.32\linewidth}
        \includegraphics[trim=6mm 2mm 7mm 2mm,clip,width=\linewidth]{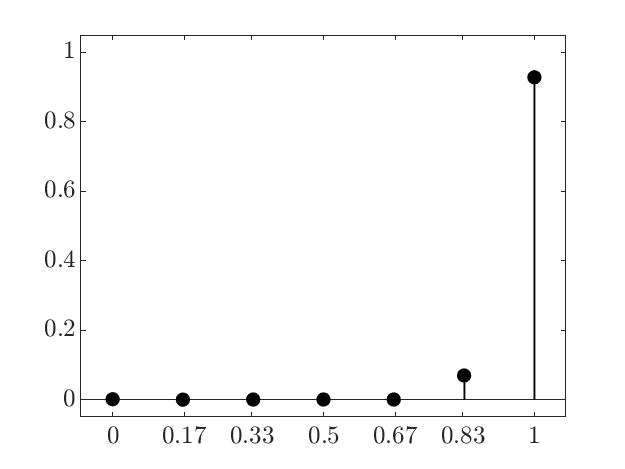}
    \end{subfigure}\\[0.3cm]

    \begin{subfigure}{0.32\linewidth}
        \includegraphics[trim=6mm 2mm 7mm 2mm,clip,width=\linewidth]{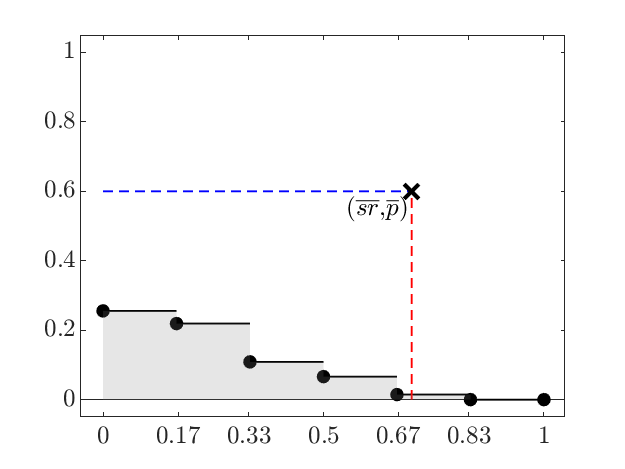}
    \end{subfigure}
    \begin{subfigure}{0.32\linewidth}
        \includegraphics[trim=6mm 2mm 7mm 2mm,clip,width=\linewidth]{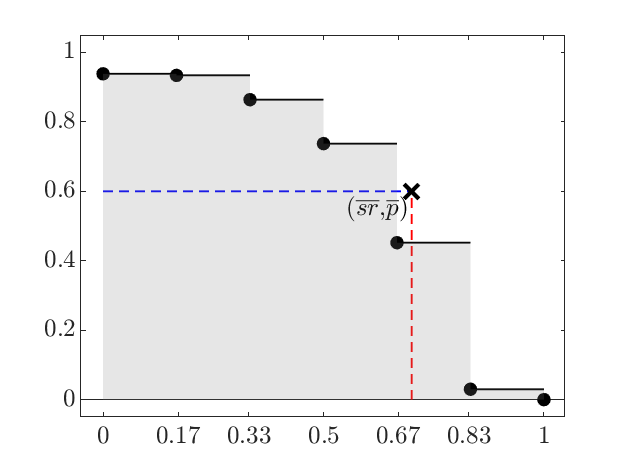}
    \end{subfigure}
    \begin{subfigure}{0.32\linewidth}
        \includegraphics[trim=6mm 2mm 7mm 2mm,clip,width=\linewidth]{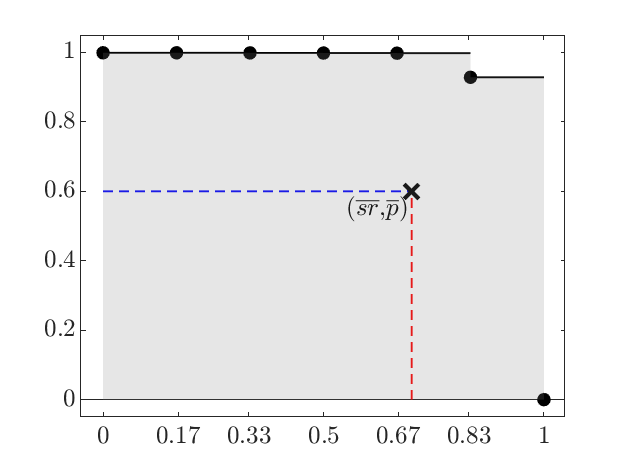}
    \end{subfigure}
    
    \caption{Probability mass function (top) and survival function (bottom) of $\RI$ for covariates $17$ (left),\, $22$ (centre) and $7$ (right). The x-axis represents the share of nonzero elements with the probabilities on the y-axis.  If we set $\overline{sr}=0.70$ (red vertical line) and $\overline{p}=0.60$ (blue horizontal line), then covariates $17$ and $22$ are to be excluded, and we include covariate $7$.}
    \label{fig:galaxy_RI}
\end{figure}

Basing the decision of covariate selection solely on thresholding the PIP could lead to misleading interpretations of results.  For this reason, we quantify the uncertainty about the decision of including the $jk$th element of $C$ through the PIP uncertainty index $\zeta_{jk}$.  The PIP uncertainty index provides a straightforward interpretation of the probabilities observed in the matrix of PIPs as a way to quantify uncertainty about the effect of each covariate on every response, which is particularly advantageous when addressing each response separately.  Notwithstanding, the decision about irrelevant covariates in the multivariate regression model is yet to be determined since some covariates may influence only a subset of the responses. For instance, covariates $1$ and $15$ have an apparent influence only on responses $\{4,5\}$ and $\{1,3\}$, respectively (Figure~\ref{fig:galaxy_C}).  Therefore, the relevance index $\RI$ is used to assess the relative importance of each covariate.  Table~\ref{tab:summaryRIn500} reports the summary statistics of the relevance index for each covariate.

\begin{table}[t!b]
\centering
\resizebox{\linewidth}{!}{%
\begin{tabular}{@{} c|cccccc c c|cccccc @{}}
\cmidrule(r){1-7} \cmidrule(l){9-15}
\multicolumn{1}{c}{$x_j$} &
  \textbf{Mode} &
  \textbf{Mean} &
  \textbf{Std} &
  \textbf{Q25} &
  \textbf{Q50} &
  \textbf{Q75} &
   &
 \multicolumn{1}{c}{$x_j$} &
  \textbf{Mode} &
  \textbf{Mean} &
  \textbf{Std} &
  \textbf{Q25} &
  \textbf{Q50} &
  \textbf{Q75} \\
\cmidrule(r){1-7} \cmidrule(l){9-15} 
\multicolumn{1}{c|}{\cellcolor{dred}1}  & \cellcolor{dred}0     & \cellcolor{dred}0.504 & \cellcolor{dred}0.392 & \cellcolor{dred}0     & \cellcolor{dred}0.667 & \cellcolor{dred}0.833 &  & \multicolumn{1}{c|}{13} & 0.833 & 0.843 & 0.049 & 0.833 & 0.833 & 0.833 \\
\multicolumn{1}{c|}{2}  & 1     & 0.986 & 0.055 & 1     & 1     & 1     &  & \multicolumn{1}{c|}{14} & 0.833 & 0.806 & 0.112 & 0.833 & 0.833 & 0.833 \\
\multicolumn{1}{c|}{3}  & 1     & 0.997 & 0.023 & 1     & 1     & 1     &  & \multicolumn{1}{c|}{\cellcolor{dred}15} & \cellcolor{dred}0     & \cellcolor{dred}0.387 & \cellcolor{dred}0.348 & \cellcolor{dred}0     & \cellcolor{dred}0.333 & \cellcolor{dred}0.667 \\
\multicolumn{1}{c|}{4}  & 1     & 0.985 & 0.047 & 1     & 1     & 1     &  & \multicolumn{1}{c|}{16} & 0.833 & 0.859 & 0.073 & 0.833 & 0.833 & 0.833 \\
\multicolumn{1}{c|}{5}  & 1     & 0.976 & 0.059 & 1     & 1     & 1     &  & \multicolumn{1}{c|}{\cellcolor{lred}17} & \cellcolor{lred}0     & \cellcolor{lred}0.111 & \cellcolor{lred}0.211 & \cellcolor{lred}0     & \cellcolor{lred}0     & \cellcolor{lred}0.167 \\
\multicolumn{1}{c|}{6}  & 1     & 0.982 & 0.052 & 1     & 1     & 1     &  & \multicolumn{1}{c|}{18} & 1     & 0.972 & 0.062 & 1     & 1     & 1     \\
\multicolumn{1}{c|}{7}  & 1     & 0.987 & 0.057 & 1     & 1     & 1     &  & \multicolumn{1}{c|}{19} & 1     & 0.96  & 0.071 & 1     & 1     & 1     \\
\multicolumn{1}{c|}{8}  & 1     & 0.985 & 0.063 & 1     & 1     & 1     &  & \multicolumn{1}{c|}{20} & 0.833 & 0.897 & 0.083 & 0.833 & 0.833 & 1     \\
\multicolumn{1}{c|}{9}  & 1     & 0.994 & 0.038 & 1     & 1     & 1     &  & \multicolumn{1}{c|}{21} & 0.833 & 0.831 & 0.068 & 0.833 & 0.833 & 0.833 \\
\multicolumn{1}{c|}{10} & 0.833 & 0.826 & 0.048 & 0.833 & 0.833 & 0.833 &  & \multicolumn{1}{c|}{\cellcolor{lred}22} & \cellcolor{lred}0.833 & \cellcolor{lred}0.659 & \cellcolor{lred}0.234 & \cellcolor{lred}0.5   & \cellcolor{lred}0.667 & \cellcolor{lred}0.833 \\
\multicolumn{1}{c|}{11} & 0     & 0     & 0.007 & 0     & 0     & 0     &  & \multicolumn{1}{c|}{23} & 0     & 0.052 & 0.147 & 0     & 0     & 0     \\
\multicolumn{1}{c|}{\cellcolor{lred}12} & \cellcolor{lred}0     & \cellcolor{lred}0.161 & \cellcolor{lred}0.233 & \cellcolor{lred}0     & \cellcolor{lred}0     & \cellcolor{lred}0.333 &  & \multicolumn{1}{c|}{}   &       &       &       &       &       &       \\ \cline{1-7} \cline{9-15} 
\end{tabular}%
}
\caption{Summary statistics of the distribution of $\RI$ in the sub-sample of $n=500$ for each covariate: mode, mean, standard deviation (Std), and the 25th, 50th and 75th quartiles (Q25, Q50, Q75). The shaded rows identify covariates with medium uncertainty ($\text{Std}(\RI_j) \geq 0.20$) and high uncertainty ($\text{Std}(\RI_j) \geq 0.30$). A description of each covariate is found in the Supplement.}
\label{tab:summaryRIn500}
\end{table}

As illustrated in the top panels of Figure~\ref{fig:galaxy_RI}, there is strong evidence for the exclusion of covariate $17$ and for the inclusion of covariate $7$. However, the exclusion or inclusion of covariate $22$ is not apparent, for there is strong variation in the distribution, and additional study should be given due consideration.\footnote{See Section 4.2 of the Supplement for the results based on alternative variable selection methods.}  The \textit{rule of thumb} in \eqref{eq:tail_RI} states not to exclude the $j$th covariate if $S_{\RI_j}(\overline{sr}) = \mathbb{P}( \RI_j > \overline{sr} ) \geq \overline{p}$, for appropriate values of $\overline{sr}$ and $\overline{p}$. By setting $\overline{sr}=0.70$ and $\overline{p}=0.60$, we are excluding covariate $17$ while maintaining covariate $7$ in the model (bottom panels of Figure~\ref{fig:galaxy_RI}), as inferred from the probability mass function of $\RI_7$ and $\RI_{22}$, accordingly.  Notice that for covariate $22$, the point $(0.70,0.60)$ lies in the rejection region for inclusion (above the curve of the tail distribution), a conclusion that would not have been reached had we required a minimum probability of $0.60$ for more than $40\%$ of the responses ($\overline{sr}=0.60$,\, $\overline{p}=0.40)$.  Under the previous specification of $\overline{sr}$ and $\overline{p}$, the covariates considered irrelevant in the majority of responses are $1,\, 11,\, 12,\, 15,\, 17,\, 22,\, 23$.


The methodology applied to the entire subset of $n=3,438$ objects reduces the percentage of zero elements in the sparse estimate of the coefficient matrix from $28\%$ to $14\%$ and the number of zero rows by $3$.  The quantity of covariates with medium and high PIP uncertainty indices decreases accordingly, in line with stronger information provided by the data to require more variables that explain the outcomes.

\section{Discussion} \label{sec:conclusion}

We have proposed BRECS, a novel Bayesian approach for estimating the rank of the matrix of coefficients along with parameters in a reduced-rank regression model.  Our method employs a mixture prior to rank estimation and shrinkage prior on the factor matrix resulting from the decomposition of the coefficient matrix for shrinking its irrelevant entries to $0$.  Furthermore, variable selection is achieved by adopting SAVS to obtain a sparse estimate of $C$, in conjunction with the uncertainty about this estimation through the relevance index. By employing our method, researchers can avoid post-processing steps and obtain a quantification of uncertainty in estimating the rank, together with accurate statistical inference for the coefficient matrix.  

The results of our simulation study suggest that RRcs exhibits superior performance over RRn in terms of converging more accurately to the true rank and being considerably faster in computational time.  We observed that the algorithm's performance deteriorates more rapidly as the number of responses increases compared to the number of covariates.  Thus, enhancing the model's scalability remains an area for future research.  Overall, our method provides a reliable estimation of the coefficient matrix, even in cases where the rank is underestimated, effectively preventing over-fitting and resulting in a satisfactory approximation of the true matrix. 

Our proposed approach was applied to real datasets about the chemical composition of tobacco leaves and photometric galaxy data.  The obtained results are consistent with the findings presented in the literature, adding a quantification of the uncertainty about the obtained estimates.  Additional domains for its applicability should be explored, including genomics \citep{hilafu2020omics} and macroeconomics \citep{Reinsel2022book}. The latter explicitly suggests extending our research to the time series model and tensor regression \citep{billio2023tensor,luo2022tensor}.  
Given the prevalence of models with incomplete response matrices in biostatistics, it is imperative to develop Bayesian approaches to address these scenarios. It is thus a promising avenue for further investigation \citep{mai2022optimal}.

\section*{Acknowledgments}
The authors gratefully acknowledge the participants at the 10th Italian  Congress of Econometrics and Empirical Economics, the 13th European Seminar on Bayesian Econometrics, the 2nd Bergamo Workshop in Econometric and Statistics, the Hi-Di-NET Final Workshop, and the Bayesian Young Statisticians Meeting (BAYSM) for their helpful feedback. This research used the Computational resources provided by the Core Facility INDACO, a project of High-Performance Computing at the University of Milan.

\section*{Supplementary Material}
The Supplement is available upon request to the authors.

\bibliographystyle{apalike}
\bibliography{references}

\end{document}